\documentclass[twocolumn,showpacs,preprintnumbers,amsmath,amssymb,superscriptaddress,prd]{revtex4}

\usepackage{rotating}
\usepackage{graphics}
\usepackage{epsfig}
\usepackage{bm}

\newcommand{\Figdir}{fig}
\newcommand{\myfiglabel}[1]{\label{#1}}

\newcommand{\psiprime}{\ensuremath{\psi '}}
\newcommand{\ccbar}{\ensuremath{c \overline{c}}}
\newcommand{\ee}{\ensuremath{e^+e^-}}
\newcommand{\lplm}{\ensuremath{l^+l^-}}
\newcommand{\mm}{\ensuremath{\mu^+\mu^-}}
\newcommand{\jpsi}{\ensuremath{J/\psi}}

\newcommand{\jpsiee}{\ensuremath{\jpsi\to e^+e^-}}
\newcommand{\jpsimm}{\ensuremath{\jpsi\to \mu^+\mu^-}}

\newcommand{\mgev}{\ensuremath{\, \mathrm{GeV}/c^2}}

\newcommand{\mmev}{\ensuremath{\, \mathrm{MeV}/c^2}}

\newcommand{\pinot}{\ensuremath{\pi^0}}

\newcommand{\Rchic}{\ensuremath{R_{\chi_c}}}
\newcommand{\chic}{\ensuremath{{\chi_{c}}}}
\newcommand{\chicz}{\ensuremath{{\chi_{c0}}}}
\newcommand{\chico}{\ensuremath{{\chi_{c1}}}}
\newcommand{\chict}{\ensuremath{{\chi_{c2}}}}
\newcommand{\chicj}{\ensuremath{{\chi_{cJ}}}}
\newcommand{\chicochict}{\ensuremath{\frac{N_{\chi_{c1}}}{N_{\chi_{c2}}}}}
\newcommand{\Rot}{\ensuremath{R_{12}}}

\newcommand{\njpsi}{\ensuremath{{N_{J/\psi}}}}
\newcommand{\nxc}{\ensuremath{{N_{\chi_{c}}}}}
\newcommand{\nxco}{\ensuremath{{N_{\chi_{c1}}}}}
\newcommand{\nxct}{\ensuremath{{N_{\chi_{c2}}}}}
\newcommand{\nxcjm}{\ensuremath{{N_{\chi_{cJ,M}}}}}
\newcommand{\njpxcjm}{\ensuremath{N_{J/\psi}^{\chi_{cJ,M}}}}

\newcommand{\gam}{\ensuremath{\gamma}}
\newcommand{\Dm}{\ensuremath{\Delta M}}

\newcommand{\epsg}{\ensuremath{\varepsilon_{\gamma}}}
\newcommand{\epsgo}{\ensuremath{\varepsilon_{\gamma}^{\chi_{c1}}}}
\newcommand{\epsgt}{\ensuremath{\varepsilon_{\gamma}^{\chi_{c2}}}}
\newcommand{\epsgj}{\ensuremath{\varepsilon_{\gamma}^{\chi_{cJ}}}}
\newcommand{\epsdir}{\ensuremath{\varepsilon_{J/\psi}^{dir}}}
\newcommand{\epso}{\ensuremath{\varepsilon_{J/\psi}^{\chi_{c1}}}}
\newcommand{\epst}{\ensuremath{\varepsilon_{J/\psi}^{\chi_{c2}}}}
\newcommand{\epsxc}{\ensuremath{\varepsilon_{J/\psi}^{\chi_c}}}
\newcommand{\epsjp}{\ensuremath{\varepsilon_{J/\psi}}}

\newcommand{\NIM}[3]       {Nucl.\ Instr.\ Methods~{\bf A#1} (#2) #3}
\newcommand{\NPA}[3]       {Nucl.\ Phys.~{\bf A#1} (#2) #3}
\newcommand{\NPB}[3]       {Nucl.\ Phys.~{\bf B#1} (#2) #3}
\newcommand{\PLB}[3]       {Phys.\ Lett.~{\bf B#1} (#2) #3}

\newcommand{\PRD}[3]       {Phys.\ Rev.~{\bf D#1} (#2) #3}
\newcommand{\PRL}[3]       {Phys.\ Rev.\ Lett.~{\bf #1} (#2) #3}

\newcommand{\ZPC}[3]       {Z.~Phys.~{\bf C#1} (#2) #3}
\newcommand{\EPJ}[3]       {Eur.\ Phys.\ J.~{\bf C#1} (#2) #3}

\newcommand{\JPG}[3]       {J.\ Phys.\ G~{\bf #1} (#2) #3}
\newcommand{\hb} {HERA-B}

\newcommand{\etal} {{\it et~al.}}

\newcommand{\institute}[1]  {}
\begin{document}

\title{ \bf
Production of the Charmonium States \chico\ and \chict\ in Proton Nucleus Interactions at \\[0.3cm] $\sqrt{s} = 41.6\,$GeV \\[0.3cm]
}
\collaboration{ \bf The \hb\ Collaboration}
\author{
I.~Abt$^{24}$,
M.~Adams$^{11}$,
M.~Agari$^{14}$,
H.~Albrecht$^{13}$,
A.~Aleksandrov$^{30}$,
V.~Amaral$^{9}$,
A.~Amorim$^{9}$,
S.~J.~Aplin$^{13}$,
V.~Aushev$^{17}$,
Y.~Bagaturia$^{13,37}$
% ...

\vspace{5mm}
\noindent
% ...
$^{9}${\it LIP Coimbra, P-3004-516 Coimbra,  Portugal~$^{f}$} \\
% ...
$^{11}${\it Institut f\"ur Physik, Universit\"at Dortmund, D-44221 Dortmund, Germany~$^{d}$} \\
% ...
$^{13}${\it DESY, D-22603 Hamburg, Germany} \\
$^{14}${\it Max-Planck-Institut f\"ur Kernphysik, D-69117 Heidelberg, Germany~$^{d}$} \\
% ...
$^{17}${\it Institute for Nuclear Research, Ukrainian Academy of Science, 03680 Kiev, Ukraine~$^{h}$} \\
% ...
$^{24}${\it Max-Planck-Institut f\"ur Physik, Werner-Heisenberg-Institut, D-80805 M\"unchen, Germany~$^{d}$} \\
% ...
$^{30}${\it DESY, D-15738 Zeuthen, Germany} \\
% ...
$^{37}${\it visitor from High Energy Physics Institute, 380086 Tbilisi, Georgia} \\
% ...
$^{d}$ supported by the Bundesministerium f\"ur Bildung und Forschung, FRG, under contract numbers 05-7BU35I, 05-7DO55P, 05-HB1HRA, 05-HB1KHA, 05-HB1PEA, 05-HB1PSA, 05-HB1VHA, 05-HB9HRA, 05-7HD15I, 05-7MP25I, 05-7SI75I \\
% ...
$^{f}$ supported by the Portuguese Funda\c c\~ao para a Ci\^encia e Tecnologia under the program POCTI \\
% ...
$^{h}$ supported by the National Academy of Science and the Ministry of Education and Science of Ukraine \\
% ...
 
}
\noaffiliation

\begin{abstract}
\vspace{3mm}
A measurement of the ratio $\Rchic = (\chic \to \jpsi + \gam)/
\jpsi$ in pC, pTi and pW interactions at $920$ GeV/c
($\sqrt{s}=41.6$ GeV) in the Feynman-x range $-0.35 < x_F^{\jpsi} < 0.15$ is
presented. Both \mm\ and \ee\ \jpsi\ decay channels are observed
with an overall statistics of about 15000 \chic\ events, which is by
far the largest available sample in pA collisions. The result
is $\Rchic = 0.188\pm0.013_{st}{^{+0.024}_{-0.022}}_{sys}$ averaged over the
different materials, when no \jpsi\ and \chic\ polarisations are considered.
The \chico\ to \chict\ production ratio $\Rot = {\Rchic}_1/{\Rchic}_2$ is
measured to be $1.02\pm0.40$, leading to a cross section 
ratio $\frac{\sigma(\chico)}{\sigma(\chict)} =0.57\pm0.23$. The dependence of
\Rchic\ on the Feynman-x of the \jpsi, $x_F^{\jpsi}$, and its transverse
momentum, $p_T^{\jpsi}$, is studied, as well as
its dependence on the atomic number, A, of the target.
For the first time, an extensive study of possible biases on \Rchic\ and
\Rot\ due to the dependence of acceptance on the polarization states of \jpsi\
and \chic\ is performed.
By varying the polarisation parameter, $\lambda_{obs}$, of all produced \jpsi's
by two sigma around the value measured by \hb, and considering
the maximum variation due to the possible \chico\ and \chict\ polarisations,
it is shown that \Rchic\ could change by a factor between 1.02 and 1.21 and 
\Rot\ by a factor between 0.89 and 1.16.  
\end{abstract}
%
\pacs{ \\
{13.20.Gd}{~Decays of \jpsi, $\Upsilon$ and other quarkonia }\\
{13.85.-t}{~Hadron-induced high- and super-high-energy interactions }\\
{24.85.+p}{~Quarks, gluons, and QCD in nuclei and nuclear processes }\\
{13.88.+e}{~Polarisation in interactions and scattering }\\
} 
\maketitle
\pagebreak

\section{Introduction}\label{intro}
Since the discovery of charmonium more than thirty years ago, its production 
in hadronic collisions has attracted considerable theoretical and experimental 
interest for a variety of reasons. In particular the question of the 
production mechanism, which 
requires an understanding of the hadronisation process in the non-perturbative
regime and, in addition, the influence of nuclear matter, are of particular
importance since the suppression of \jpsi\ production has been considered
as a possible indicator of the quark-gluon plasma~\cite{QGP}. 

The theoretical treatment of quarkonium production is usually broken into two 
steps: the creation of a heavy
quark pair in interactions of the colliding partons, calculable by means
of perturbative QCD, and the transition to a bound state, involving poorly
understood non-perturbative processes and even more problematic nuclear 
effects. A variety of approaches have been developed to describe 
quarkonium production such as the Color Evaporation Model (CEM)~\cite{CEM}, 
the Color Singlet Model (CSM)~\cite{CSM}, and non-relativistic QCD 
(NRQCD)~\cite{NRQCD}. A measurement of the fraction of \jpsi\ coming 
from the decay of other charmonium states (feed-down) provides 
useful tests of the model predictions. While a 
rather rich sample of data on \jpsi\ production exists, the available data
on the production rates, or even the experimentally simpler fractional 
production rates, of the other charmonium states suffer from imprecision. 
Moreover very little experimental information is available on the 
possible polarisation of the produced charmonium states.

In this paper we report on the production of the charmonium states \chico\ and 
\chict\ in collisions of a 920 GeV proton beam with nuclear targets. The 
\chic\ mesons are identified via their radiative decay into \jpsi\ mesons which
in turn are decaying into lepton pairs. The production and decay chain is:
\begin{equation}
        p\,A \to \chi_{c} + X ; \ \ \ \ \chi_{c} \to \gamma\,  J/\psi \to \gamma\, l^+ l^- \ \ \ (l=e,\, \mu ).
\end{equation}
To minimise systematic uncertainties the \chic\ rates are normalised to the 
total production rate of \jpsi. We define \Rchic, the fraction of \jpsi\ 
originating from radiative \chic\ decays:
\begin{equation}
\Rchic = \frac{ \sum_{i=1}^{2} 
\sigma({\chic}_{i}) Br({\chic}_{i} \rightarrow \jpsi \gam)}{\sigma(\jpsi)}
\label{eq:intro:rxc}
\end{equation}
where $Br({\chic}_{i} \rightarrow \jpsi \gam)$ are the branching ratios for
the different ${\chic}_{i} \to \jpsi \gam$ decays, $\sigma({\chic}_{i})$ are 
their production cross sections per nucleon and $\sigma(\jpsi)$ is the total
\jpsi\ production cross section per nucleon. In Tab.~\ref{tab:intro:properties}
the main properties of the three \chic\ states (\chicz, \chico\ and \chict) are
reported. Due to the negligible $\chicz \to \jpsi\gam$ branching ratio, we 
limit our study to \chico\ and \chict\ production.
\begin{table}[h]
\begin{center}
\begin{tabular}{c c c c c}
\hline
\hline
State & Mass ($MeV/c^2$) & Width ($MeV/c^2$) & BR($\to \jpsi\gam$) \\
\hline
\chicz\ & $3414.76\pm0.35$ & $10.4\pm0.7$ & $(1.30\pm0.11)\%$ \\
\chico\ & $3510.66\pm0.07$ & $0.89\pm0.05$ & $(35.6\pm1.9)\%$ \\
\chict\ & $3556.20\pm0.09$ & $2.06\pm0.12$ & $(20.2\pm1.0)\%$ \\
\hline
\hline
\end{tabular}
\caption{\it Properties of the three \chic\ states~\cite{PDG}.}
\label{tab:intro:properties} 
\end{center}
\end{table}

In Section~\ref{production} an overview of the physics motivations for our 
measurement is given together with a survey of the existing experimental 
results. In Section~\ref{realdata} the experiment and the
data taking conditions are described and in Section~\ref{method} the
principle of the measurement is explained. The simulation used for evaluating 
detection efficiencies and the measurements in
the muon and electron decay channels are described in 
Sections~\ref{montecarlo},~\ref{Jpsi_count} and~\ref{chic_count}, respectively. In 
Section~\ref{polar}, the effect of possible \jpsi\ and \chic\ polarisations 
(not directly measured in this analysis) on \Rchic\ and 
$\Rot={\Rchic}_1/{\Rchic}_2$ is discussed. A discussion of systematic 
uncertainties 
and the final result are given in Sections~\ref{systematics} and~\ref{results},
respectively.

\section{$\chi_c$ Production}\label{production}
\subsection{QCD Models of Charmonium Production}\label{models}

In the Color Evaporation Model (CEM), charmonium production is described as 
the creation of $c \bar c$ pairs with an invariant mass below the $D \bar D$ 
threshold. Their hadronisation is mediated by the emission of soft gluons 
which do not significantly alter the kinematics of the $c\bar c$ system.
The production rates of the various charmonium states are predicted to be
proportional to each other and 
independent of the projectile, target and energy. Most experiments 
measuring \Rchic\ in proton and pion induced 
interactions~\cite{cobb}-\cite{E706} provide compatible results
with the predicted value $\Rchic \approx 0.4$~\cite{ramonavogt}. 
The assumption of the universality of charmonium hadronisation implies that 
the value of \Rchic\ should be independent of the kinematic variables  
$x_F$ and $p_T$ of the produced charmonium state~\cite{ramonavogt} ($x_F$ is 
the Feynman
variable in the nucleon-nucleon center-of-mass system; $p_T$ is the transverse 
momentum relative to the incoming beam).

In the Color Singlet Model (CSM), the quark pair is created in a hard 
scattering reaction as a colour 
singlet (CS) with the same quantum numbers as the final quarkonium. Since 
two gluons can form a colourless C-even state such as the \chic\ states, but 
at least three gluons are needed to form a colourless C-odd state such as the 
$\psi$ 
states, $\psi$ production in this model is suppressed by an additional factor 
$\alpha_s$. As a result, the \jpsi\ production rate should be dominated by 
feed-down from radiative \chic\ decays and \Rchic\ is predicted to be close to 
1. Most of the proton induced \chic\ measurements are in disagreement with this
assumption~\cite{cobb}-\cite{xcherab}.

In response to the disagreement between the CSM and measurements in most
charmonium production features~\cite{cdf-jpsi}, a more generalised 
perturbative QCD approach for charmonium production, 
`non-relativistic QCD' (NRQCD), was developed which includes not only
\ccbar\ pairs produced as colour singlets but also as colour octets (CO). The 
CO states subsequently evolve into the observed charmonium by soft gluon 
emission. At the \hb\ beam energy of $920$ GeV, the dominant production 
process is $gg$ fusion 
which contributes both to CO and CS states. Therefore \chic\ production 
dominates the CS part of \jpsi\ production while direct \jpsi\ and 
\jpsi\ from \psiprime\ decay are produced via CO states.
The predicted ratio, $\Rchic \approx 0.3$~\cite{ramonavogt} is
in agreement with most of the existing measurements in
proton induced interactions~\cite{cobb}-\cite{xcherab}.
NRQCD predicts only small differences in the differential cross sections 
of the different charmonium states as a function of $x_F$, mostly at 
large values of $x_F$. More visible differences can arise when considering
nuclear-matter effects (A-dependence) due to the differing absorption
probabilities of the various charmonium and pre-charmonium states
in nuclei~\cite{ramonavogt}.

\subsection{Interactions with Nucleons and A-dependence}

The CEM model and NRQCD differ in their predictions of the suppression of 
the charmonium production rate per nucleon in interactions with heavy nuclei
compared to interactions with single proton targets. Suppression can occur in 
interactions of the generated $c \bar c$ quarks with nuclear matter which 
could lead to an $x_F$ dependence: for $x_F > 0$, the formation length of the 
final charmonium state exceeds the size of the nucleus, while for $x_F < 0$, 
an increasingly larger fraction is formed already inside the nucleus. In the 
context of the CEM, only one proto-charmonium state exists and thus for 
$x_F > 0$, the differences in suppression between \jpsi, \psiprime\ and the 
\chic\ states should be small. For $x_F < 0$, the \chic\ and \psiprime\ states 
should be more suppressed than the \jpsi\ due to their larger 
interaction cross sections. In the context of NRQCD, substantial differences 
in the suppression of the various charmonium states are expected even for 
$x_F > 0$ since the wave function of the CO states extends over a much larger 
distance and the resulting interaction cross section is considerably larger 
than that of the CS states~\cite{ramonavogt}. 

\subsection{\Rchic\ and the Quark-Gluon Plasma}\label{QGP}

The so-called ``anomalous'' suppression of \jpsi\ has been proposed as a
possible indicator of the formation of a quark-gluon plasma~\cite{QGP} and 
such suppression has subsequently been reported by several 
experiments~\cite{NA50}~\cite{NA60}~\cite{PHENIX}.
Nevertheless the conclusion that the reported suppression is indeed anomalous 
is contingent on the full understanding of normal suppression mechanisms, 
i.\,e.\ those existing in the absence of a quark-gluon plasma as is expected 
to be the case in proton-nucleus reactions.
In this respect the measurement of the fraction of \jpsi\ arising from 
feed-down decays (\chic\ and \psiprime) is important since the
anomalous suppression is expected to be sensitive to the mass and binding 
energy of the different charmonium states. Directly produced \jpsi\ survive 
in the quark-gluon plasma up to about $1.5\, T_c$~\cite{Nardi}, $T_c$ being 
the critical 
temperature, while \chic\ and \psiprime\ states dissociate just above $T_c$. 
Thus several drops in the distribution of charmonium survival 
probability as a function of the temperature are expected, with the size of 
the drops dependant on the fractions \Rchic and $R_{\psiprime}$ 
($R_{\psiprime} = \frac{ \sigma(\psiprime) Br(\psiprime \rightarrow \jpsi X)}
{\sigma(\jpsi)}$). 
Experimentally, 
only the first drop has been reported~\cite{NA50}~\cite{NA60}~\cite{PHENIX}, 
and is interpreted as indicating the disassociation of \chic\ and \psiprime.
Several models attempt to describe the totality of experimental data on \jpsi\
suppression. They generally assume $\Rchic \sim 0.3$ and 
$R_{\psiprime} \sim 0.1$.
Nevertheless all the proposed models fail to simultaneously describe all the 
existing data, as they all overestimate the suppression unless other
effects, such as \jpsi\ regeneration, are assumed to describe the RHIC 
data~\cite{Nardi}.
From the value of \Rchic\ shown in Sect.~\ref{results} and the result
from~\cite{hbpsiprime}, $R_{\psiprime}   \approx 7\%$,
$\Rchic+R_{\psiprime} \approx 0.27$, which is lower than generally 
assumed.

\subsection{Previous Measurements}\label{measurements}

The production of \chic\ has been measured both in proton- and pion-induced 
reactions on various nuclear targets and in $pp$ and $p \bar p$ 
interactions~\cite{cobb}-\cite{E706}. Tab.~\ref{tab:production:exper-list} 
lists all the published measurements of \chic\ production in hadronic 
interactions and reports their most relevant features. From this table some 
observations can be made:
\begin{itemize}
\item all fixed target measurements are based on at most a few hundreds \chic;
\item all experiments observe only one of the two \jpsi\ decay channels 
(\ee\ or \mm);
\item the photon efficiency never exceeds $30\%$;
\item most measurements are performed in the positive $x_F$ range.
\end{itemize}
Tab.~\ref{tab:production:exper-res} shows the measured values of \Rchic\ 
and/or of the \chic\ cross sections separately for proton and pion induced 
reactions. The values shown in Figs.~\ref{fig:production:rchic-proton} 
and~\ref{fig:production:rchic-pion} have been updated
using the current PDG values~\cite{PDG} for the \chic\ and \jpsi\ decay 
branching ratios, and the \jpsi\ cross sections obtained 
from~\cite{ourcharmonium}.

The available data scatter strongly, well beyond their respective
uncertainties, and no energy dependence is discernible.
The proton data seem to favour a value 
$\Rchic  \sim 0.3$, supporting the prediction of NRQCD, but the quality of
the available data does not allow a firm conclusion.

\begin{table*}
\begin{small}
\begin{tabular}{c c c c c c c c c c c c}
\hline
\hline
Exp. & beam/ & $\sqrt(s)$ & \lplm\ & \gam\  & \epsg\ & $x_F$ & $p_T$ & $E_{\gam}$ cut &\njpsi\ &\nxc\ &$\chi_{ci}$\\
   & target & GeV & & det. & $\%$ & & GeV/c & GeV & & & sep.\\
\hline
ISR~\cite{cobb}&pp&$<55>$& \ee\ &d& & & &$>0.4$&$658$&$31\pm11$&n\\
R702~\cite{clark}&pp&52.4,62.7& \ee\ &d& & &$<3$&0.4-0.6&$975$& & n\\
ISR~\cite{kourkoumelis}&pp&62& \ee\ &d& & &$<5$&$>0.4$& & & n\\
E610~\cite{bauer}&pBe&19.4,21.7& \mm\ &d&$16$&0.1-0.7&$<2$&3-50&$157\pm17$&$11.8\pm5.4$&f\\
E705~\cite{Antoniazzi}&pLi&23.8& \mm\ &d&27&-0.1-0.5&0.-0.4&$>1.0$&$6090\pm90$&$250\pm35$&f\\
E771~\cite{Aleksopoulos}&pSi&38.8& \ee\ &c&0.8&$>0.0$& &0.25-0.7&$11660\pm139$&66&y\\
\hb\ \cite{xcherab}&pC,Ti&41.6& $\{^{\ee}_{\mm}$ &d&30&-0.25-0.15& &$E_T>$1.0&$4420\pm100$&$370\pm74$&n\\
\hline
CDF~\cite{cdf1},\cite{cdf2} &$p \bar p$&1800& \mm\ &$\{^c_d$ & $\{^{}_{15}$& &$>4.0$&$>1.0$&$\{^{88000}_{32642\pm185}$&$\{^{119\pm14}_{1230\pm72}$&$\{^y_n$\\
\hline
\hline
E369~\cite{E369}&$\pi^-Be,p$&20.2& \mm\ &d& &0-0.8&$<3$&$<5$&160&$17.2\pm6.6$&n\\
WA11~\cite{WA11}& $\pi^-Be$ & 18.7 & \mm\ & c & 1 & & & & 44750 & 157 & y\\
IHEP140~\cite{IHEP}&$\pi^-p$&8.6& \ee\ &d& &$>0.4$&$<2$&$>2$&120&10&n\\
E673~\cite{E673}&$\pi^-Be$&20.6& \mm\ &d&21& & &10-25&$1056\pm36$&$84\pm15$&n\\
E610~\cite{E610}&$\pi^-Be$&18.9& \mm\ &d&19&0.1-0.7&$<2$ &3-50&$908\pm41$&$53.6\pm17.1$&f\\
E705~\cite{E705}&$\{^{\pi^-}_{\pi^+}-Li$&23.8& \mm\ &d&27& & & &$\{^{5560\pm90}_{12470\pm160}$&$\{^{300\pm35}_{590\pm50}$&n\\
E672/706~\cite{E706}&$\pi^-Be$&31.1& \mm\ &$\{^d_c$&$\{^{11}_2$&0.1-0.8& &$>10$&$7750\pm110$&$\{^{379\pm66}_{105\pm18}$&$\{^f_y$ \\
\hline
\hline
\end{tabular}
\caption{\it Previous \Rchic\ measurements in hadronic collisions. Symbols: 
\gam\ detection (d=direct, c=\gam-conversion). \chico-\chict\ separation 
(y=yes, n=no, f=with 2-states fit).}
\label{tab:production:exper-list}
\end{small}
\end{table*}

\begin{table*}
\begin{small}
\begin{center}
\begin{tabular}{c c c c c c c c c}
\hline
\hline
Exp. & \multicolumn{4}{c|}{Measured values} & \multicolumn{4}{c}{Updated values} \\
\hline
&$R\chi_c$&$\frac{\sigma(\chico)}{\sigma(\chict)}$&$\sigma(\chico)$&$\sigma(\chict)$&$R\chi_c$&$\frac{\sigma(\chico)}{\sigma(\chict)}$&$\sigma(\chico)$&$\sigma(\chict)$\\
 & & &(nb/n)&(nb/n)& & &(nb/n)&(nb/n)\\
\hline 
\cite{cobb}&$0.43\pm0.21$& & & & $0.43\pm0.21$ & & & \\
\cite{clark}&$0.15^{+0.10}_{-0.15}$& & & & $0.15^{+0.10}_{-0.15}$& & & \\
\cite{kourkoumelis}  & 0.47(8)          & & & &0.47(8) & & & \\
\cite{bauer}&0.47(23)&0.24(28)&64(81)&268(136)&0.47(23)&0.24(28)&39(49)&162(81)\\
\cite{Antoniazzi}&0.30(4)&0.08(25)(15)&31(62)(3)&364(124)(36)&0.30(4)&0.09(29)(17)&24(48)(2)&244(83)(16)\\
\cite{Aleksopoulos}&0.77(30)(15)&0.53(20)(7)&526(138)(64)&996(286)(134)&0.76(29)(16)&0.61(24)(4)&488(128)(56)&805(231)(92)\\
\cite{xcherab}       &0.32(6)(4)& & & &0.32(6)(4)& & &\\
\hline
\cite{cdf1}&0.297(17)(57)&1.04(29)(12)& & &0.297(17)(57)&1.19(33)(14)& &\\
\hline
\hline
\cite{E369}& 0.70(28)& & & &0.70(28)& & &\\
\cite{WA11}&0.30(5)&0.68(28)&65(18)&96(29)&0.30(5)&0.79(28)&58(13)&74(19)\\
\cite{IHEP}&0.44(16)&1(fix)&28(10)&28(10)&0.44(16)&1(fix)&22(8)&22(8)\\
\cite{E673}& 0.37(9)& 1.12(42)& & &0.37(9)&1.11(41)& &\\
\cite{E610}&0.31(10)&0.96(64)&130(56)&134(64)&0.31(10)&0.98(74)&102(43)&104(49)\\
\cite{E705}&0.40(4)& & & &0.40(4)& & & \\
\cite{E705}&0.37(3)&0.70(15)&131(17)&189(31)&0.37(3)&0.80(16)&101(13)&126(19)\\
\cite{E706}&0.443(41)(35)&0.57(16)&464(87)&815(168)&0.443(41)(35)&0.65(18)&356(66)&544(107)\\
\hline
\hline
\end{tabular}
\caption{\it \Rchic, $\frac{\sigma(\chico)}{\sigma(\chict)}$, $\sigma(\chico)$ 
and $\sigma(\chict)$ results in hadronic collisions.
Statistical and systematic uncertainties are shown in brackets (less 
significant digits). See text for an explanation of the updated values.}
\label{tab:production:exper-res}
\end{center}
\end{small}
\end{table*}
\begin{figure*}[htb]
\centering
\resizebox{1.0\textwidth}{!}{%
\epsfig{file=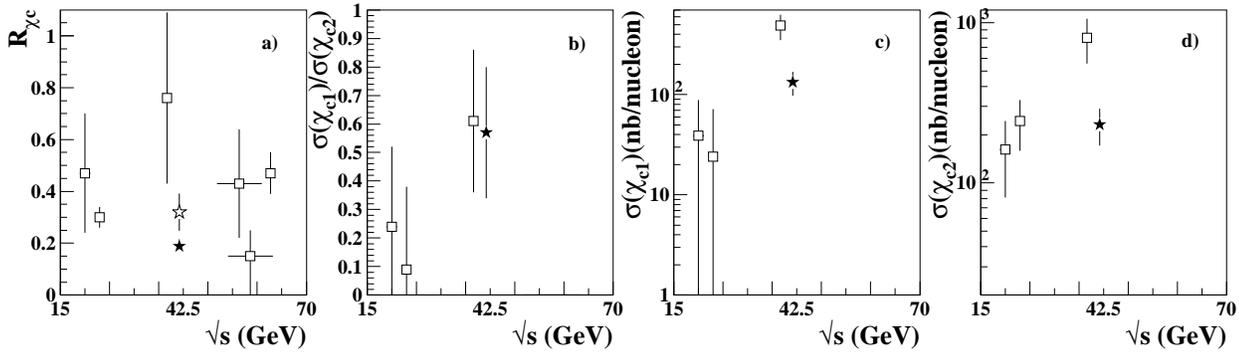}}
\caption{\it Summary of experimental results on \Rchic\ (a),
$\frac{\sigma(\chico)}{\sigma(\chict)}$ (b), $\sigma(\chico)$ (c) and
$\sigma(\chict)$ (d) in pN interactions. \hb\ results: open star (old),
full star (new).}
\myfiglabel{fig:production:rchic-proton}
\end{figure*}
\begin{figure*}[htb]
\centering
\resizebox{1.0\textwidth}{!}{%
\epsfig{file=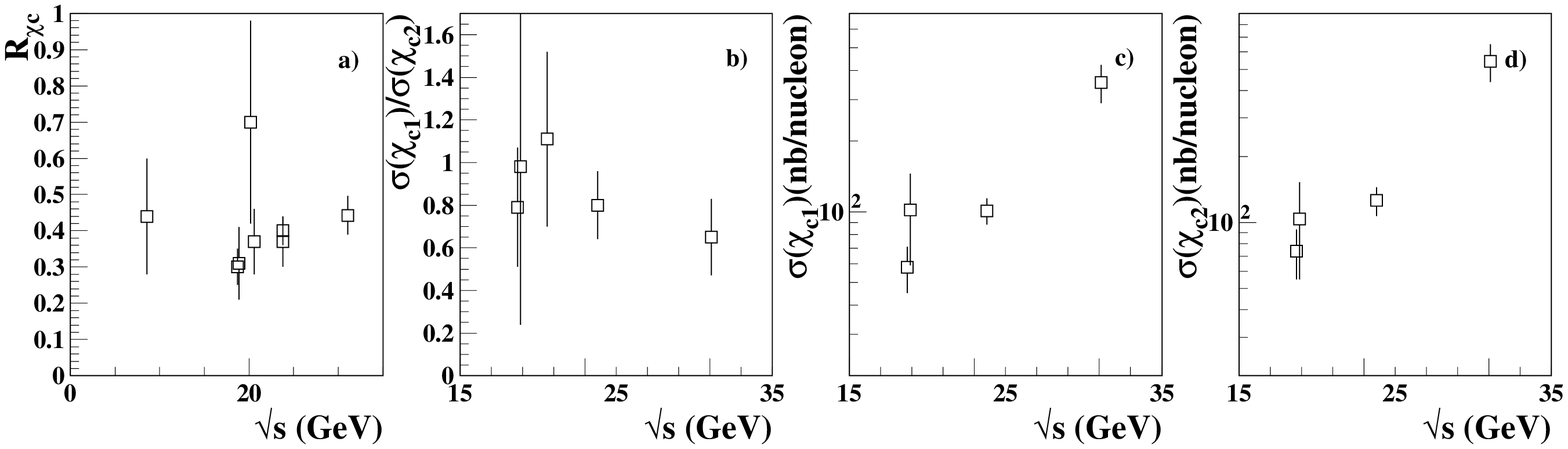}}
\caption{\it Summary of experimental results on \Rchic\ (a),
$\frac{\sigma(\chico)}{\sigma(\chict)}$ (b), $\sigma(\chico)$ (c) and 
$\sigma(\chict)$ (d) in $\pi N$ interactions.}
\myfiglabel{fig:production:rchic-pion}
\end{figure*}

\section{The Experiment and the Data Sample}\label{realdata}
The \hb\ detector~\cite{HB_tdr} was a forward magnetic spectrometer used to 
study the interactions of the 920 GeV proton beam ($\sqrt s = 41.6$ GeV) of 
the HERA accelerator on a variety of nuclear targets. The detector components 
relevant for this analysis are the wire target system~\cite{target} which 
could be dynamically positioned in the halo of the proton beam, the Silicon 
Vertex Detector (VDS)~\cite{vds}, the dipole magnet of 2.13 Tm, the drift-tube
Tracking System (OTR)~\cite{otr}, the Ring Imaging Cherenkov Counter 
(RICH)~\cite{RICH}, the sampling Electromagnetic Calorimeter 
(ECAL)~\cite{ECAL} and the Muon Detector (MUON)~\cite{MUON}. 

The data sample of about 160 million events used for this analysis was 
acquired at an interaction rate of about 5 MHz with a dedicated di-lepton 
trigger~\cite{HB_bbar} 
in order to select both \jpsiee\ and \jpsimm\ 
final states. In total about 300,000 \jpsi\ were reconstructed, distributed
almost equally in the two decay channels. Nine different wire configurations
were used, both in single and double wire runs. The wire materials used were 
carbon (C, $\approx$64\% of the full statistics), tungsten (W, $\approx$31\%),
and titanium (Ti, $\approx$5\%). Continuous online monitoring
ensured stable running conditions, and further offline data quality
checks were applied to select only runs with properly functioning
detector and trigger components.

\section{Experimental Method}\label{method}
\Rchic\ is defined in Eq.~\ref{eq:intro:rxc}. The quantity to be measured is:
\begin{equation}\label{eq:method:rxc}
\Rchic =\frac{\frac{\nxco \cdot \epsdir}{\epso \cdot \epsgo} + \frac{\nxct \cdot \epsdir}{\epst \cdot \epsgt}}{ \njpsi + \frac{\nxco}{\epsgo} \cdot (\frac{\epsdir}{\epso}-1)+ \frac{\nxct}{\epsgt} \cdot (\frac{\epsdir}{\epst}-1)}
\end{equation}
where \njpsi\ is the total number of observed \jpsi 's, \nxco\ (\nxct) is 
the number of counted \chico 's (\chict 's), \epsdir\ is the direct \jpsi\ 
total detection efficiency, including trigger losses, reconstruction and cut 
selection, \epso\ (\epst) is the total detection efficiency for \jpsi\ coming 
from \chico\ (\chict) decay and \epsgo\ (\epsgt) 
is the identification efficiency of the photon from \chico\ (\chict) decay 
for events with identified \jpsi 's. The measurement method consists of 
evaluating \njpsi\ by analysis of the di-lepton invariant mass spectra and 
\nxco, \nxct\ by analysis of the \jpsi-\gam\ invariant mass spectra, for 
events with selected \jpsi\ candidates. The efficiency terms in 
Eq.~\ref{eq:method:rxc} are extracted from the MC simulation.

The production ratio of the two states can be determined using:
\begin{equation}
\Rot = \frac{{\Rchic}_1}{{\Rchic}_2}=\frac{\nxco}{\nxct} \cdot \frac{\epst 
\cdot \epsgt}{\epso \cdot \epsgo} 
\label{eq:method:R12}
\end{equation}
(where ${\Rchic}_1+{\Rchic}_2 = \Rchic$) and the production cross section 
ratio can be evaluated using:
\begin{equation}
\frac{\sigma({\chic}_{1})}{\sigma({\chic}_{2})} = \Rot
\frac{Br({\chic}_2 \rightarrow \jpsi \gam)}
{Br({\chic}_1 \rightarrow \jpsi \gam)}
\label{eq:method:sigma12}
\end{equation}

In order to perform an internally consistent analysis, the same
procedure and cuts are applied to both the \ee\ and \mm\ channels except for 
the lepton particle identification (PID) requirements.

\subsection{$J/\psi$ Selection}\label{jpsi}

Leptons from \jpsi\ decay are selected from the triggered tracks, re-fitted 
using offline alignment constants and taking into account multiple 
scattering when extrapolating to the target. A $\chi^2$ probability of the 
track fit $>0.3\%$ is required. Additional PID cuts are applied depending on 
the lepton channel. 

In the muon channel, a muon likelihood is constructed from information in the 
MUON detector and is required to be greater than $5\%$ and a kaon likelihood 
is constructed from RICH information and required to be less than $99\%$. 

In the electron channel a more complex set of PID cuts is needed. 
First the calorimeter is searched for a cluster consistent with having been 
caused by a Bremsstrahlung photon emitted in front of the magnet~\cite{ECAL}. 
Since the presence of such a cluster very effectively identifies electrons, 
the cut values used for the remaining two particle identification criteria can 
be substantially relaxed when such a Bremsstrahlung cluster is found. The 
additional two criteria are a more restrictive matching requirement between 
the OTR track of the electron candidate and its corresponding ECAL cluster, 
and a requirement that the track momentum be consistent with the deposited 
calorimeter energy.

Once opposite sign lepton candidates (\mm\ or \ee) are selected, their 
common vertex is fitted and the $\chi^2$ probability of the fit is required 
to be greater than $1\%$. In a few percent of the events, more than one 
di-lepton combination pass all cuts, in which case only the one with 
the lowest product of track-fit $\chi^2$ is retained. Finally, the invariant 
mass of the di-lepton pair is calculated and required 
to be within $2\sigma$ of the nominal $J/\psi$ mass, with $\sigma = 36$ 
MeV/$c^2$ in the muon channel and $64$ MeV/$c^2$ in the electron channel. 

\subsection{Mass Difference Plot}\label{massdiff}

The next step after \jpsi\ selection is the identification of suitable photon 
candidates. A photon is defined as a reconstructed ECAL cluster~\cite{villa}
with at least three contiguous hit cells. 
The cluster energy, $E^{\gamma}$, is required to be at least 0.3 GeV and the 
cluster transverse energy, $E_T^{\gamma}$, is required to be at least 0.2 GeV,
for an optimal cluster reconstruction. Furthermore, the ECAL cell with the 
highest 
energy deposit of the cluster is required to contain at least 80\% of the 
total cluster energy in order to provide some discrimination against showering 
hadrons. Clusters which match reconstructed tracks are excluded
unless the matching track is formed only from hits behind the 
magnet and point to the selected di-lepton vertex. Such tracks are mainly 
from conversions of event-related photons behind the magnet. Finally, because 
of high background near the proton beam pipe, clusters in an elliptic region 
around the pipe ($\sqrt{x_{clust}^2/4 + y_{clust}^2} < 22$~cm, where 
$x_{clust}$ and $y_{clust}$ are the horizontal and vertical positions of the 
cluster with respect to the beam) are excluded.  

Since a photon from a \chic\ decay cannot be distinguished from the others 
in the event (on average $\sim 20$), the combinatorial background to the 
\chic\ signal is very large, as will be shown in Sect.~\ref{bck}.

To largely eliminate the uncertainty due to di-lepton mass resolution, the 
analysis is performed using the mass difference, $\Dm = M(\jpsi\gam)-M(\jpsi)$.
The dominant contribution to the mass difference resolution is the intrinsic 
photon energy resolution determined by the ECAL.

\subsection{Background Description}\label{bck}

The analysis crucially depends on the background shape being correctly 
described. We distinguish between ``physical'' backgrounds (due to 
the decay of heavier states which include a \jpsi\ and one or more photons 
in their decay products) and ``combinatorial" background (due to photons from 
the event combined with di-leptons which share no parent resonance).
The combinatorial background by far dominates. The only significant physical 
background comes from $\psi(2S)\to J/\psi\pi^0\pi^0$ which contributes at the 
level of $\approx 15\%$ of the $\chi_c$ rate but with a rather flat  
distribution in the \Dm\ spectrum. The shape of this 
background is estimated from Monte Carlo and subtracted after proper 
normalisation.

A ``Mixed Event'' (ME) procedure is adopted for modelling the combinatorial 
background: a \jpsi\ candidate from one event (``event-A'') is mixed with the 
photons of several ($\approx 20$) other selected events (which we all call
``event-B''). 
Event-B is required to have the same neutral cluster multiplicity as event-A 
to ensure similar photon energy spectra. Furthermore, the angular difference 
between the vector sums of transverse momenta of all photons in event-A and 
event-B is required to be no more than $2\pi/20$ to ensure the events to be 
kinematically similar and thus to have similar acceptance. 

Extensive tests, both with Monte Carlo and the data itself, were performed to 
verify the ME procedure. For example, using the data, the combination of 
photons with \lplm\ pairs in the \jpsi\ side bands (defined as the di-lepton 
mass intervals outside $3\sigma$ of the nominal $J/\psi$ mass, see 
Sect.~\ref{jpsi}) in the SE (``same event'') and with \lplm\ pairs 
inside the \jpsi\ mass window in the ME spectra show no reflection of the
\chic\ peak and the 
SE over ME ratio for these events is found to be flat. The normalisation of 
the ME spectrum is incorporated into the fit of the \Dm\ spectrum as a free 
parameter (see Sect.~\ref{chic_count}).

\section{The Monte Carlo simulation}\label{montecarlo}
\subsection{Event generator and detector simulation}\label{generator}

In the HERA-B Monte Carlo, the basic process $pN \to Q \bar Q X$ is simulated,
first, by generating the heavy quarks ($Q \bar Q$), including hadronisation, 
with PYTHIA 5.7~\cite{pythia}; secondly, the energy of the remaining part of 
the process (X) is given as an input to FRITIOF~\cite{FRITIOF}, which is
used to simulate the interactions inside the nucleus. PYTHIA describes by 
default the charmonium 
production based on the Color Singlet Model. Further colour singlet 
and colour octet processes were therefore added, according to the NRQCD 
approach~\cite{xcherab}. Differing kinematic distributions for directly
produced \jpsi\ and \jpsi\ from feed-down decays generated according to this 
model result in slightly different acceptances: $\sim 78.3\%$ for direct 
\jpsi\ and $\sim 77.6\%$ for \jpsi\ from both \chico\ and \chict, with no 
significant difference between the two \chic\ states.

In the simulation, both direct \jpsi\ and \chic\ states are generated with no 
polarisation and all results are given under this assumption. The effects of 
\jpsi\ and \chic\ polarisation are discussed and treated separately 
(see Sect.~\ref{polar}).

The detector response is simulated using GEANT 3.21~\cite{GEANT} and includes
individual detector channel resolutions, noise, efficiencies and calibration 
precision. The second level trigger algorithm is applied to the simulated 
detector hits and the first level trigger efficiency is taken from an 
efficiency map obtained from the data itself. The generated Monte Carlo is 
reconstructed with the same package used for reconstructing the data and the 
same analysis cuts are applied to the MC and the data.

In order to check the MC material description, which influences the 
photon efficiency determination, three different studies were performed by 
using the Bremsstrahlung tag~\cite{ECAL}, the \pinot\ signal (where
the decay photons are seen as neutral clusters or as converted photons), and
the converted photons (see Sect.~\ref{eff} and~\ref{systematics}). 

The predicted resolution of the \chico\ and \chict\ states is found to
be $\sim 0.032~\mgev$, in agreement with real data (see Sect.~\ref{chic_count}).

\subsection{\jpsi\ and photon efficiency}\label{eff}

According to Eq.~\ref{eq:method:rxc}, the ratios of efficiencies for \jpsi\ 
coming from the decay of the \chic\ states to that of directly produced \jpsi\ 
are needed. These ratios are estimated from MC and the values obtained are 
reported in Table \ref{tab:montecarlo:effjp}. As can be seen from the table, 
these ratios are independent of target, decay channel and \chic\ state, within 
the errors. 

The photon detection efficiencies are also evaluated with the MC although an 
additional correction factor derived from the data was found to be needed, as 
will be discussed below. For the efficiency evaluation, the same analysis  
as for the data is performed, but the photon from the \chic\ decay is 
selected using MC generation information and checked for acceptance after all 
cuts are applied. The alternative of extracting the number of \chic 's from 
the MC using the ME background subtraction applied to the data, and thus 
inferring the photon efficiency without recourse to the MC generation 
information, was found to give a compatible efficiency, but with lower 
precision. 

The MC estimate for photon detection efficiency was checked by comparing the 
efficiency derived from MC to that, obtained from data, for the detection 
by the ECAL of reconstructed electrons or positrons from photon conversions 
before the magnet. Since the average di-lepton triggered data run contains 
several thousands of such reconstructible conversions, the method affords a 
detailed check of the stability of photon detection efficiency over the run as 
well as a check of the MC.\\
The tracks from the converted photons are required to share a common VDS 
track segment and to have hits in the OTR chamber immediately before the ECAL 
(to discriminate against electrons which start to shower before the ECAL). 
When using the positron from such a pair as a probe, the electron (``tag'') is
also required to have an associated ECAL cluster with a deposited energy 
compatible with the electron track momentum (and vice versa). For selected 
electron and positron probes, the ECAL is searched for a geometrically 
matching cluster and the ratio of the deposited ECAL energy to the track 
momentum (``E/p ratio") is entered into a histogram. Signal to background 
ratios of the order of 15 are achieved. The E/p ratio histogram of the probe 
as well as the corresponding E/p histogram of the tag are fitted to gaussians 
to describe the signal and third order polynomials for the background 
description. The fit describes the data well with $\chi^2$ values typically 
equal to or less than the number of degrees of freedom. The efficiency is 
extracted from the fit parameters. The ratio of MC efficiencies estimates to 
the efficiency derived by this method is found to be $1.144 \pm 0.034$, with
the quoted uncertainty dominated by run to run variations. Roughly half the 
difference between efficiency estimates from MC and data can be attributed 
to a higher ECAL cluster multiplicity in the data compared to the MC -- when a 
cluster caused by a photon from a \chic\ overlaps with another cluster the 
photon's measured energy becomes too large and the mass estimate incorrect. 
The remaining ($\approx 7\%$) discrepancy is not understood but is likely due 
to cases where the energy deposited by the photon (electron) is considerably 
less than would be expected from gaussian statistics. 

In Tab.~\ref{tab:montecarlo:effxc} the values of efficiency and the width of
\chico\ and \chict\ are reported for the two lepton channels and for the
different target materials. 
\begin{table}[htb]
\begin{center}
\begin{tabular}{c c c c c} 
\hline
\hline
Mat. & \multicolumn{2}{c|}{ \mm } & \multicolumn{2}{c}{ \ee } \\
\hline 
  & $\frac{\epsdir}{\epso}$  & $\frac{\epsdir}{\epst}$ & $\frac{\epsdir}{\epso}$  & $\frac{\epsdir}{\epst}$  \\
\hline
C  & $0.972 (7)$ & $0.965 (5)$ & $0.970 (12)$ & $0.950 (7)$ \\
W  & $0.957 (8)$ & $0.974 (6)$ & $0.985 (14)$ & $0.955 (9)$\\ 
Ti & $1.008 (26)$ & $0.957 (17)$ & - & - \\
\hline
\hline
\end{tabular}
\caption{\it The ratio of efficiencies for detection of directly produced 
\jpsi 's to that of \jpsi 's  coming from \chico\ and \chict\ decay. The 
efficiencies for \mm\ and \ee\ channels and for each target material are given 
separately.}
\label{tab:montecarlo:effjp}
\end{center}
\end{table}
\begin{table}[h]
\begin{center}
\begin{tabular}{c c c c c}
\hline
\hline
Mat. & \epsgo\ ($\%$)& $\sigma_{\chico}~(MeV/c^2)$ & \epsgt\ ($\%$) &  $\sigma_{
\chict}~(MeV/c^2)$\\
\hline
C  & $40.5\pm0.4$ & $30.2\pm0.4$ & $41.2\pm0.2$ & $33.0\pm0.2$\\
W  & $37.1\pm0.5$ & $32.0\pm0.7$ & $38.2\pm0.3$ & $34.3\pm0.5$\\
Ti & $41.3\pm1.3$ & $31.6\pm0.8$ & $41.4\pm0.8$ & $30.2\pm0.3$\\
\hline
\hline
C  & $39.6\pm0.6$ & $32.1\pm0.4$ & $40.4\pm0.3$ & $33.0\pm0.2$\\
W  & $38.6\pm1.0$ & $33.7\pm0.8$ & $38.3\pm0.6$ & $35.4\pm0.4$\\
\hline
\hline
\end{tabular}
\caption{\it Photon detection efficiencies and the expected widths of 
$\Delta M$ peaks for \chico\ and \chict\ for the muon (first part) and the 
electron (second part) channels. }
\label{tab:montecarlo:effxc}
\end{center}
\end{table}

\section{Event counting}\label{event_count}
\subsection{$J/\psi$ counting}\label{Jpsi_count}
\begin{figure*}[htb]
\centering
\resizebox{1.0\textwidth}{!}{%
\epsfig{file=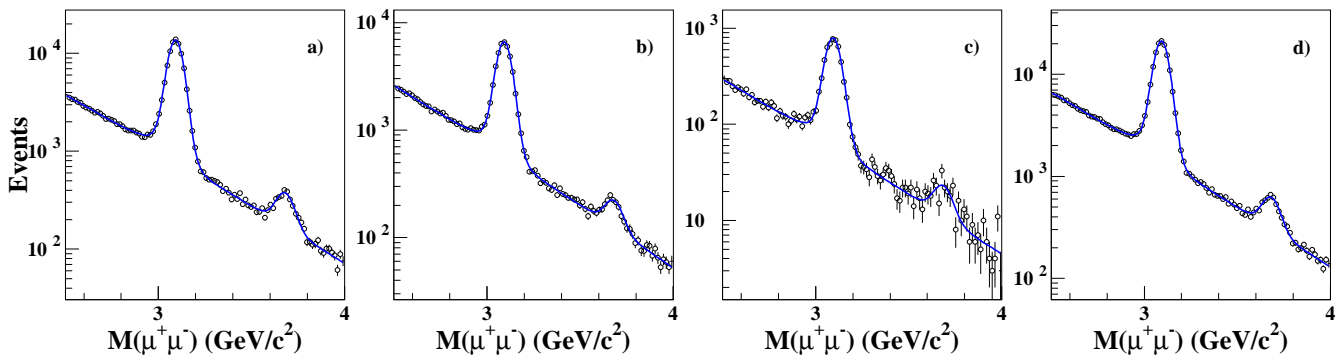}}
\caption{\it \mm\ invariant mass spectra in the muon channel for C (a), W (b), 
Ti (c) and full sample (d). The bin width is 15\mmev.}
\label{fig:muons:jpsi}
\end{figure*}

\paragraph{The muon channel:} 
The \mm\ invariant mass spectra for C, Ti, and W samples as well as the summed 
spectrum are shown in Fig.~\ref{fig:muons:jpsi}  along with a fitted curve. 
The fit includes the \jpsi\ and \psiprime\  peaks, each described by a 
superposition of three gaussians with a common mean plus a radiative tail to 
describe the photon emission process 
$\psi\to\mu^+\mu^-\gamma$~\cite{hbpsiprime}, and an exponential to describe 
the background. The numbers of \jpsi\ within the mass window used for \chic\ 
selection are reported in Tab.~\ref{tab:muons:njpsi}.
 
\begin{figure*}[htb]
\centering
\resizebox{1.0\textwidth}{!}{%
\epsfig{file=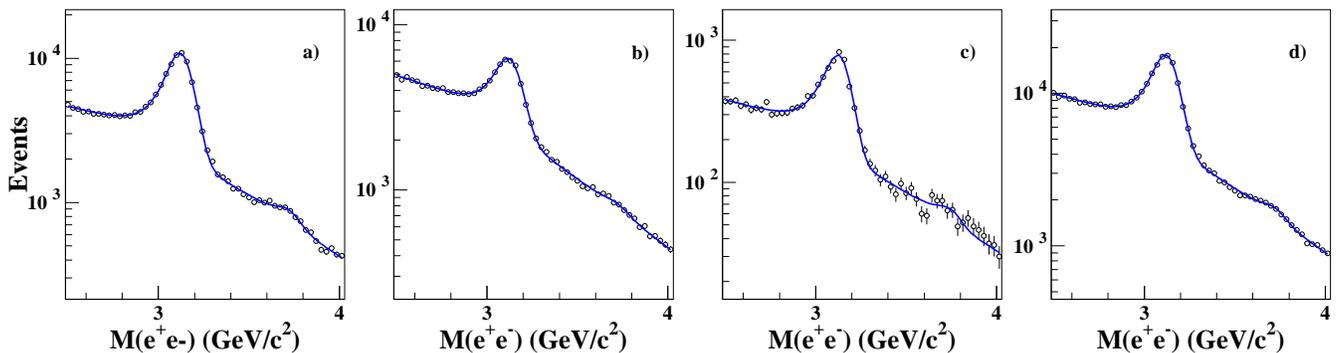}}
\caption{\it \ee\ invariant mass spectra for C (a), W (b), Ti (c)
 and full sample (d). The bin width is 30 \mmev.}
\label{fig:ele:jpsi}
\end{figure*}

\paragraph{The electron channel:} 
The \ee\ invariant mass spectra for the different materials and the
full sample are shown in Fig.~\ref{fig:ele:jpsi}. The fit used for the
signals (\jpsi\ and \psiprime) includes a Gaussian for the right part of the 
peaks and, for the left part, a Breit-Wigner to take into account the 
Bremsstrahlung tail, while a gaussian (exponential) describes the background
in the low (high) mass region with the requirement of continuity of the 
functions and of the first derivatives. The numbers of \jpsi\ within the mass 
window used for \chic\ selection are reported in Tab.~\ref{tab:muons:njpsi}.

\begin{figure*}
\centering
\resizebox{1.0\textwidth}{!}{%
\epsfig{file=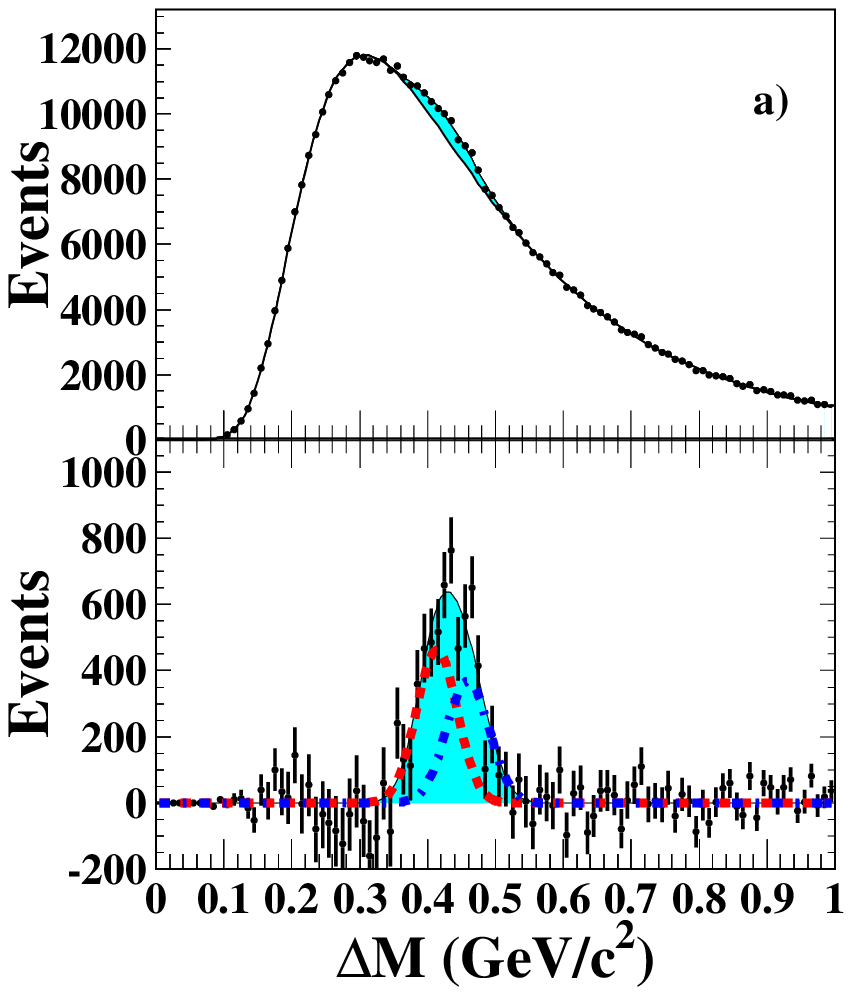}
\epsfig{file=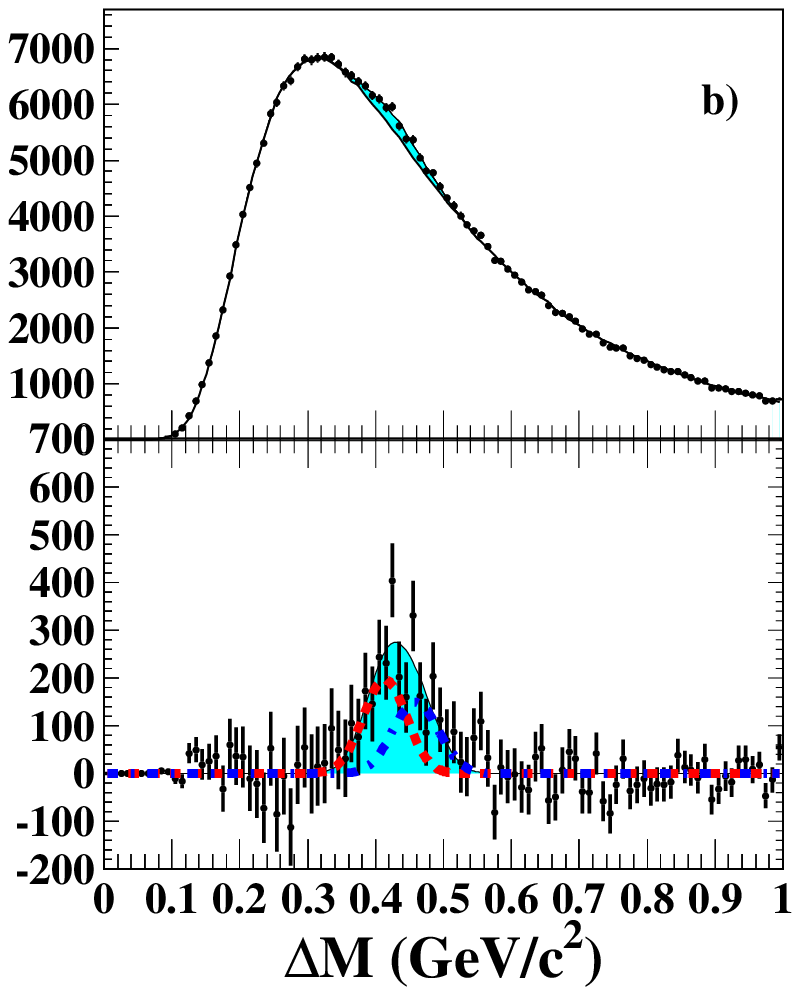}
\epsfig{file=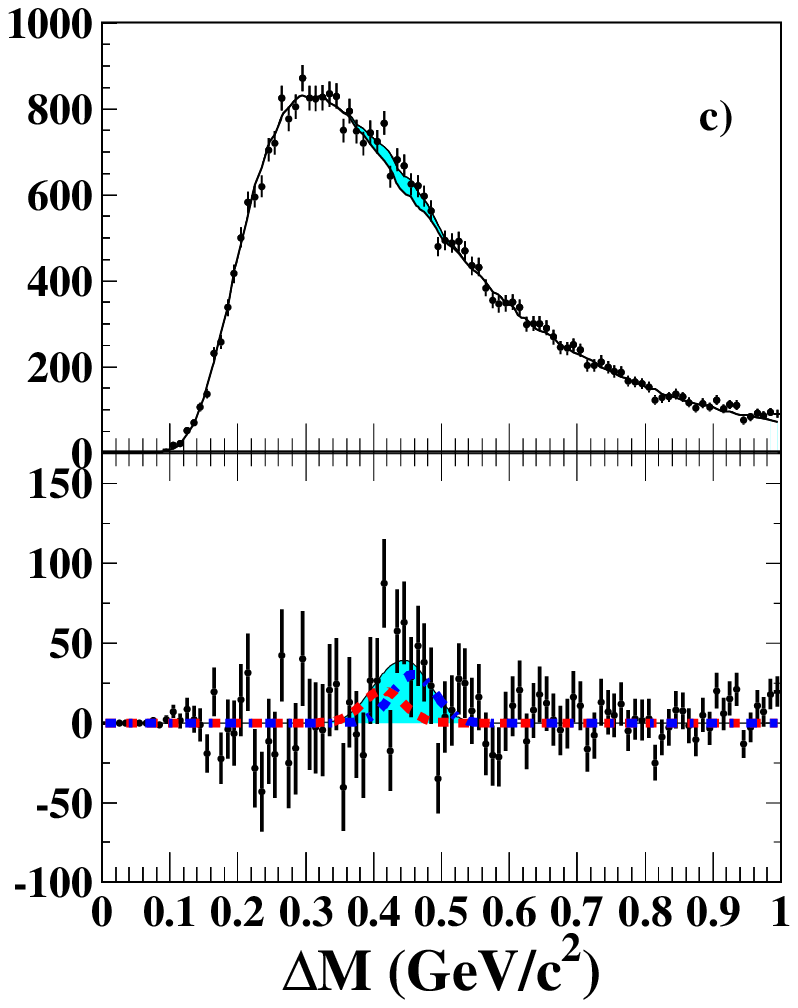}
\epsfig{file=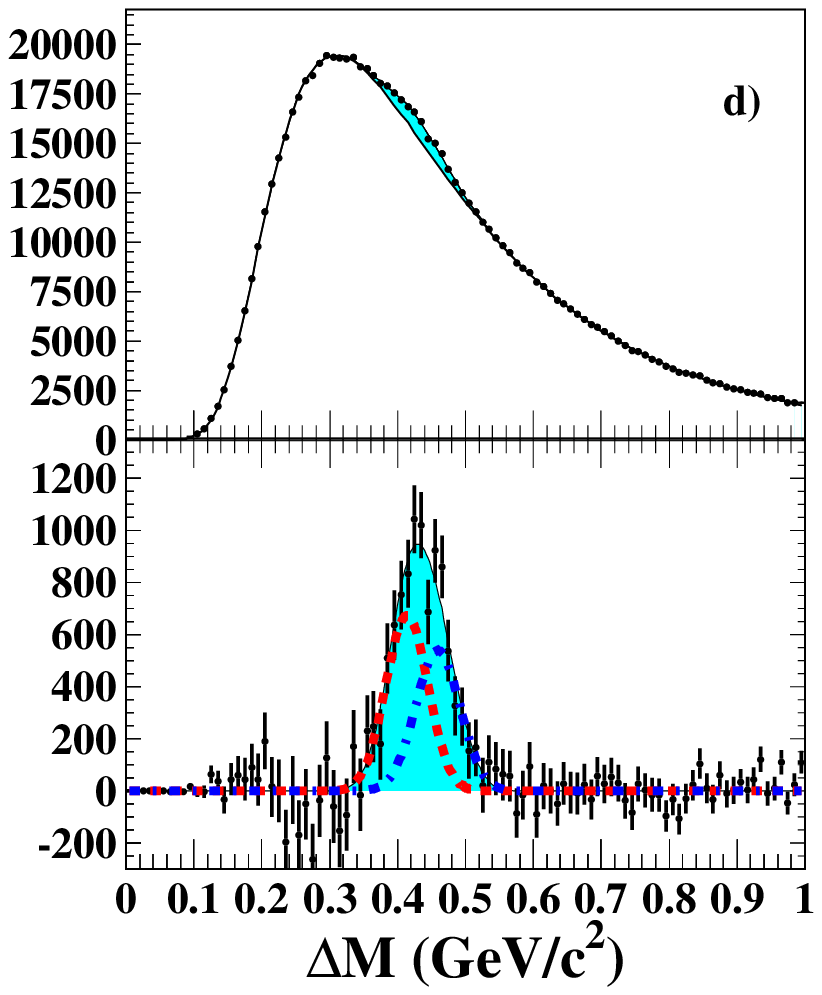}
}
\caption{\it \Dm\ spectra in the muon channel for C (a), W (b), Ti (c) and 
full sample (d). The bin width is 10\mmev. In the background subtracted spectra
the broken curves are the fitted \chico\ and \chict\ states.}
\label{fig:muons:xc}
\end{figure*}
\begin{figure*}
\centering
\resizebox{1.0\textwidth}{!}{%
\epsfig{file=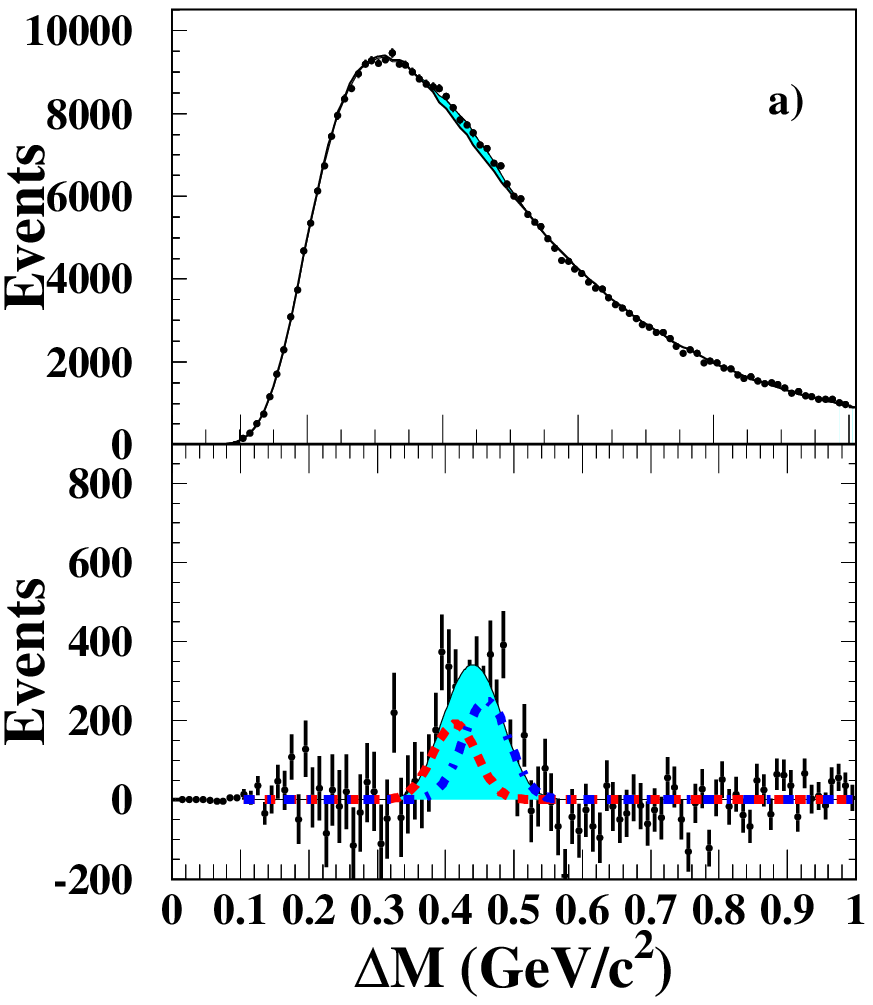}
\epsfig{file=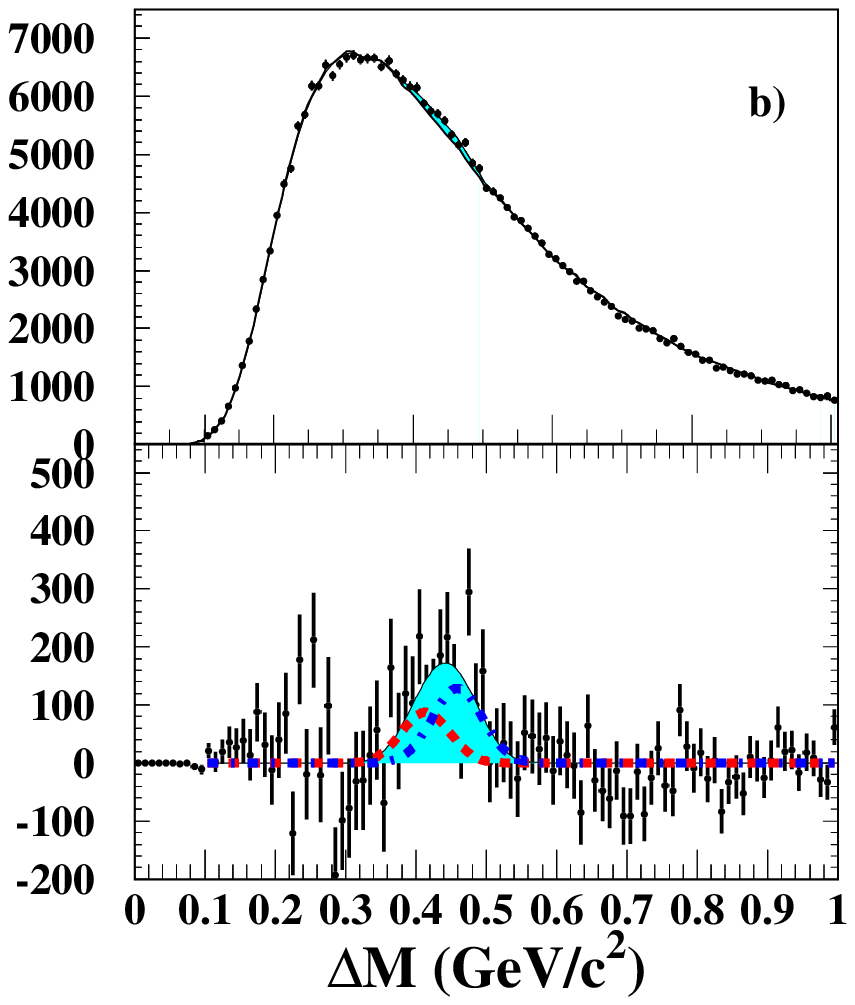}
\epsfig{file=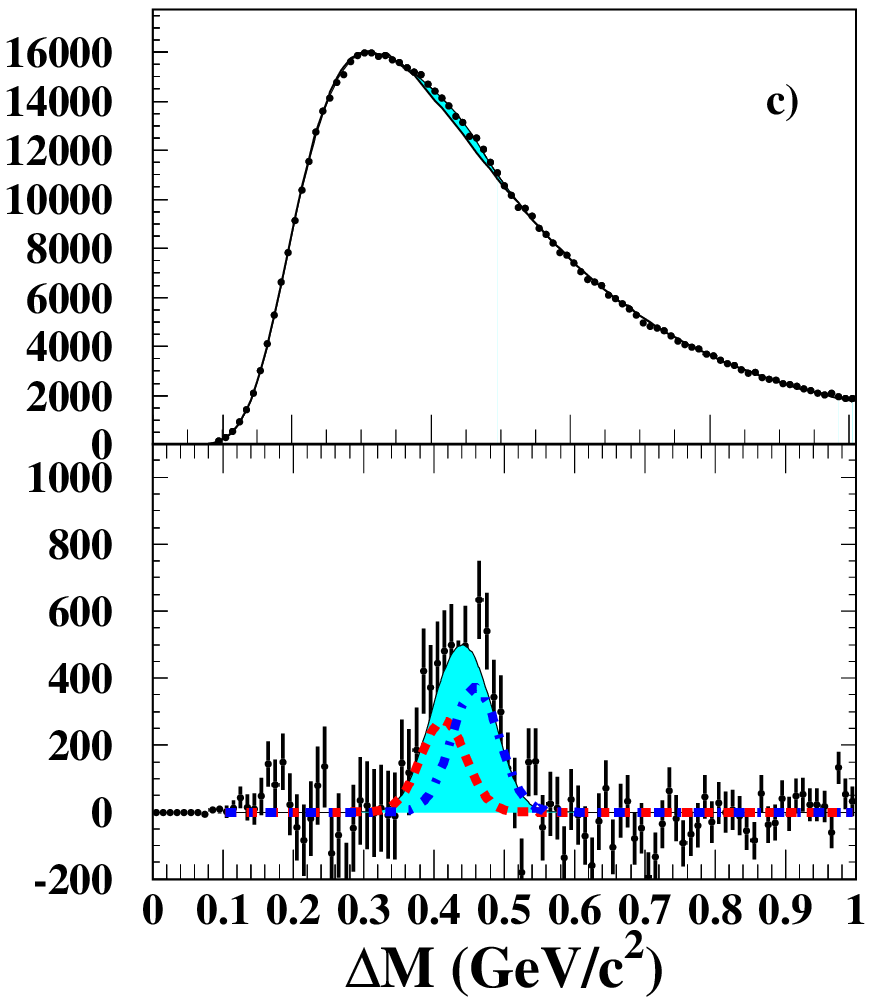}
}
\caption{\it \Dm\ spectra in the electron channel for C (a), W (b) and 
full sample (c). The Ti spectrum is not shown due to the low statistics. 
The bin width is 10 \mmev. In the background subtracted spectra
the broken curves are the fitted \chico\ and \chict\ states.}
\label{fig:ele:xc}
\end{figure*}

\subsection{$\chi_c$ counting}\label{chic_count}
The \Dm\ spectra are shown in Fig.~\ref{fig:muons:xc} for the muon channel, 
and Fig.~\ref{fig:ele:xc} for the electron channel. In the upper parts of 
these figures, the SE data are indicated by points. Fits to the ME and SE 
samples are shown as solid lines.  The two curves are not distinguishable 
except in the \Dm\ region between 0.3 and 0.6 \mgev\ where the ME curve is 
below the SE curve. The fit to the SE spectrum uses the ME parameterization to
describe the background and Gaussian distributions for the signal, as
described below. In order to evaluate the quality of the background 
description, the background subtracted spectra are shown below the fitted \Dm\ 
spectra for visual representation only. 
A clear \chic\ signal is visible both in the carbon and tungsten samples, 
while in the titanium sample, the significance of the \chic\ signal is at the 
level of $\sim3.5\sigma$ only.

The detector resolution for the two \chic\ states is 
comparable with their mass difference, resulting in a single \chic\ peak in 
the \Dm\ spectrum. It is nevertheless possible to count separately the number 
of \chico\ and \chict\ (and therefore measure \Rot) by using a fit with two 
gaussians with some of the parameters fixed, namely: 
\begin{enumerate}
\item $\Dm_{\chico} = 0.4137~\mgev$~\cite{PDG};
\item $\sigma_{\chico}$ fixed according to the MC prediction from 
Tab~\ref{tab:montecarlo:effxc};
\item $\Dm(\chict)-\Dm(\chico) = 0.0455~\mgev$~\cite{PDG};
\item $\frac{\sigma_{\chict}}{\sigma_{\chico}} = 1.05$ as predicted by MC.
\end{enumerate}
The free parameters are $N_{\chic }=N_{\chico +\chict}$, 
\chicochict\ and the ME normalisation parameter. In Tab.~\ref{tab:muons:nxc} 
the values of the fitted $N_{\chic }$ and \chicochict\
are reported for both the electron and muon
channels together with the \chic\ counting obtained with a single-gaussian fit
where the \chic\ is considered as a single peak. 

In order to verify the assumptions made, a systematic study of the effect of 
releasing the different fixed parameters or varying them within a range around 
the assumed values was done, as well as a cross check of the \chic\ counting 
with the signal modelled as a single gaussian to describe both \chic\ states. 
The results of these studies are discussed in Sect.~\ref{systematics}.
\begin{table}
\begin{center}
\begin{tabular}{c c c c c} 
\hline
\hline
Mat. & \multicolumn{2}{c|}{ \mm } & \multicolumn{2}{c}{ \ee } \\ \hline
 & $N_{\jpsi}$ & $\sigma_{\jpsi}$ & $N_{\jpsi}$ & $\sigma_{\jpsi}$ \\
 & & (MeV/$c^2$) & & (MeV/$c^2$) \\
\hline
   C   &  $80400\pm300$ & $35.6\pm0.2$ & $50030\pm530$  & $64.2\pm0.8$\\ 
   W   &  $47750\pm200$ & $36.0\pm0.3$ & $23460\pm480$  & $66.1\pm1.6$\\ 
   Ti  &  $ 4700\pm 70$ & $37.1\pm0.7$ &  $3530\pm150$  & $58.8\pm2.8$ \\ 
   Tot & $122900\pm400$ & $35.8\pm0.1$ & $77020\pm700$  & $64.3\pm0.8$\\ 
\hline
\hline
\end{tabular}
\end{center}
\caption{\it Total numbers of $J/\psi$ events per target material and for
the full data set in the \mm\ and \ee channels.}
\label{tab:muons:njpsi}
\end{table}
\begin{table}[htb]
\begin{center}
\begin{tabular}{c c c c c}
\hline
\hline
Mat. & \multicolumn{1}{c|}{1-G fit} & \multicolumn{3}{c}{2-G fit} \\ 
\hline
 \multicolumn{5}{c}{ \mm } \\ 
\hline
 & $N_{\chic}$ & $N_{\chico+\chict}$ & \chicochict & $\sigma(\chico)$ \\
 & & & & (GeV/$c^2)$ \\ 
\hline
 C  & $6280\pm510$ & $6390\pm420$ & $1.20\pm0.26$ & $0.030$ \\
 W  & $3120\pm560$ & $2830\pm330$ & $1.26\pm0.52$ & $0.032$ \\
 Ti & $ 390\pm110$ & $ 390\pm110$ & $0.63\pm0.63$ & $0.030$ \\
 Tot& $9570\pm710$ & $9630\pm550$ & $1.19\pm0.24$ & $0.031$ \\
\hline
\hline
 \multicolumn{5}{c}{ \ee } \\ 
\hline
 & $N_{\chic}$ & $N_{\chico+\chict}$ & \chicochict & $\sigma(\chico)$ \\
 & & & & (GeV/$c^2)$ \\ 
\hline
C   & $3890\pm480$ & $3600\pm390$ & $0.79\pm0.31$ & $0.032$ \\
W   & $2080\pm370$ & $1870\pm330$ & $0.71\pm0.48$ & $0.034$ \\
Tot & $5630\pm660$ & $5250\pm500$ & $0.76\pm0.28$ & $0.033$ \\
\hline
\hline
\end{tabular}
\caption{\it Results of the fit of the \chic\ signal in the \mm\ and \ee
channels. See text for the meaning of the different fit procedures.}
\label{tab:muons:nxc}
\end{center}
\end{table}

\section{Polarisation}\label{polar}
The experimental determination of polarisation can be used to probe assumptions
on the impact of specific QCD processes and the influence of nuclear effects. 
The data available for this analysis of \chic\ production does not allow a 
determination of the polarisation of the \chic\ states because of large 
backgrounds. In the following we discuss angular distributions for the decay 
products of \chic\ and directly produced \jpsi\ states with the goal of 
investigating the possible influences of polarisation on the acceptances and 
thus on the determination of the \chic\ rates.

\subsection{\chic\ polarisation}\label{polarisation:chi_c}

The full angular distribution of final state particles in the radiative decay
\begin{equation}
\label{eq_chidecay}
    \chi_{cJ} \to \gamma J/\psi \to \gamma \ l^+l^-
\end{equation}
can be found for pure \chic\ polarisation states
$|J,M\rangle$ with $J=1,2$ and $|M|=0,...,J$ in the appendix.
The  angular distribution formulae are independent of the choice of a
particular polarisation axis (e.\ g.\ Gottfried-Jackson, Collins-Soper or 
other systems can be used). Possible coherent mixtures are not considered here 
because we assume that a study of the pure states will be sufficient to 
determine systematic acceptance effects due to polarisation.

If one assumes no azimuthal dependence for the production process, the \chic\ 
decay depends on three angles which are chosen as follows: a polar decay angle,
$\theta$, defining the direction of the \jpsi\ in the \chic\ rest system with 
respect to the polarisation direction; a polar angle, $\theta'$, defining the 
direction of the positive lepton in the \jpsi\ rest system  with respect to 
the \jpsi\ direction (in the \chic\ rest system); an azimuth angle, $\phi'$, 
which is the angle between the plane defined by the polarisation axis and the 
\jpsi\ direction, and the decay plane of the \jpsi. 

For a state $|J,M\rangle$ the angular distribution can be decomposed into 
terms with trigonometric expressions $T_i^J(\theta,\, \theta',\, \phi')$ and 
coefficients $K_i^{J,M}$ \cite{Olsson}:
\begin{equation}
    \label{eq_w_t_k}
    W^{J,M}(\theta,\, \theta',\, \phi')=
    \sum_i K_i^{J,M}\ T_i^{J}(\theta,\, \theta',\, \phi').
\end{equation}
The angular functions  $T_i^J(\theta,\, \theta',\, \phi')$ and the coefficients $K_i^{J,M}$, expressed in terms of helicity amplitudes, are reported in Table \ref{tab_w_t_k_j} in the appendix. With the additional assumption that for both \chic\ states only the leading multipole, the electric dipole,  contributes to the radiative decay, the coefficients $K_i^{J,M}$ are uniquely defined (see appendix for the numerical values).  The assumption that higher order multipoles can be neglected is well justified by experimental results \cite{PDG}. The pure \chic\ polarisation states are thus unambiguously defined.

\subsection{\jpsi\ polarisation}\label{polarisation:jpsi}

\subsubsection{\jpsi\ angular distributions}
In leptonic \jpsi\ decays, the \jpsi\ polarisation can be determined from the 
angular distribution of the leptons. After integrating over the azimuthal 
orientation of the decay plane of the \jpsi\ (or assuming azimuthal symmetry) 
the distribution of the polar decay angle $\theta'$ can be parameterised as:
\begin{equation}
    \label{eq_lambda}
    \frac{1}{N}\frac{dN}{d \cos\theta'}= a(\lambda)(1+\lambda \cos^{2}\theta'),
 a(\lambda)=\frac{1}{2 (1+ \lambda/3)},
\end{equation}
where $\theta'$ is the angle between the $l^{+}$ and the quantisation axis. 
The form of the distribution (\ref{eq_lambda}) is independent of the chosen 
quantisation axis (in general however the value of $\lambda$ is dependent on 
this choice).

\subsubsection{\jpsi\ polarisation measurement}
\label{polarisation:jpsi:measurement}

A measurement of \jpsi\ polarisation by the HERA-B collaboration is reported 
in~\cite{roberto_jpsi}. However the \jpsi\ sample used for this study includes 
not only directly produced \jpsi\, but also \jpsi\ from \chic\, and it is not 
possible to distinguish between the two contributions. Therefore the 
$\lambda$-value derived from the observed distribution,  $\lambda_{obs}$, has 
to be considered as the average \jpsi\ polarisation parameter, independent of 
the origin. 

The polarisation parameters have been determined using as quantisation axis,
the bisector of the angle between $\mathbf{p}_{b}$ and $\mathbf{p}_{t}$, 
where $\mathbf{p}_{b}$, $\mathbf{p}_{t}$ are  the momenta of the beam proton 
and the target nucleon, respectively, in the \jpsi\ centre-of-mass system 
(`Collins--Soper frame'). The experimental value of the polarisation 
parameter, averaged over the muon and the electron decay channels and on
the target materials, and assuming 
no dependence on $p_T^{\jpsi}$ and $x_F^{\jpsi}$ in the \hb\ acceptance, is 
\cite{roberto_jpsi}:
\begin{equation}
         \lambda_{obs}=-0.35\pm0.04.
\label{eq:lambda_obs}
\end{equation}

\subsection{Method for the evaluation of systematic uncertainties due to 
polarisation}\label{polarisation:jpsi:description}

In this section we explain the method used to estimate the systematic 
uncertainties for \Rchic\ and \Rot\ arising from possible polarisations of 
the \chic\ and directly produced \jpsi\ states. The only experimental 
constraint which can be used for these estimations is the measured  
$\lambda_{obs}$ (section \ref{polarisation:jpsi:measurement}).

\subsubsection{Principle of the method}

The efficiencies entering in the formulae for  \Rchic\ and \Rot\ 
(\ref{eq:method:rxc}, \ref{eq:method:R12}) depend in general on the 
polarisation of the \chic\ and the directly produced \jpsi\ states. The 
efficiencies will be evaluated for the \chic\ pure polarisation states 
described in section \ref{polarisation:chi_c} which will then be used to limit 
the ranges of possible \Rchic\ and \Rot\ values which will in turn be used to 
determine the uncertainties of these values in 
section~\ref{systematics:polarisation}. As in the \jpsi\ 
analysis~\cite{roberto_jpsi}, we evaluate the polarisation states in the 
Collins--Soper frame. 
Despite this specific choice and the restriction to pure polarisation states,
we assume to get an estimate of uncertainties induced by polarisation.

The formulae for  \Rchic\ and \Rot\ require the detection efficency for direct 
\jpsi\ production \epsdir\ which depends on the polarisation parameter 
$\lambda_{dir}$. Since the observed polarisation, $\lambda_{obs}$, also 
includes the effect of possible \chic\ 
polarisation, $\lambda_{dir}$ has to be disentangled from it 
using the values $\lambda_1$ and $\lambda_2$ obtained for the assumed
polarisation states of \chico\ and \chict\, respectively. 
This is done with an iterative procedure in which the yet-to-be-determined 
values of \Rchic\ and \Rot\ are used as inputs.

\subsubsection{Determination of $\lambda_{dir},\ \lambda_1,\ \lambda_2$}

Starting from Eq.~(\ref{eq_lambda}) the observed polar decay angle 
distribution can be decomposed into contributions from  directly 
produced \jpsi\ and \jpsi\ from \chico\ and \chict\ events: 
\begin{equation}
    \label{eq_lambda_mix}
   a_{obs} (1+\lambda_{obs}\cos^2\theta') = \sum_{i=dir, 1, 2} f_i\, a_i\  (1 + \lambda_i\cos^2\theta'),  
\end{equation}
with $a_i = a (\lambda_i)$.
The fractions $f_i ({i=dir, 1, 2})$ of the different types of \jpsi\ 
are determined by  \Rchic\ and \Rot . With $\sum_i f_i =1$,  $f_{1} + f_{2} = \Rchic$ and ${f_{1}}/{f_{2}}=\Rot$ one obtains
\begin{equation}
    \label{eq_lambda_note_fi}
        f_{dir}  =   1-\Rchic ,\ \
        f_{1}  =  \frac{\Rchic \Rot}{1+ \Rot}, \ \
        f_{2}  =  \frac{\Rchic }{1+ \Rot} .
\end{equation}

Since there is no direct measurement of $\lambda_{1}$ and
$\lambda_{2}$, the angular distributions corresponding to the different pure 
polarisation states $|J, M\rangle$ of \chico\ and \chict\ described in 
Sect.~\ref{polarisation:chi_c} are used to determine the
sub-ranges allowed for $\lambda_{1}$ and $\lambda_2$, out of the full
$[-1.0, 1.0]$ interval. For this purpose, the unpolarised \chic\ angular 
distribution 
is re-weighted with the corresponding function (\ref{eq_w_t_k}). The resulting 
$\cos \theta'$ distribution is then fitted with the function (\ref{eq_lambda})
which then  yields $\lambda_{1}$ or $\lambda_{2}$ corresponding to the tested 
pure \chic\ polarisation state.

Solving (\ref{eq_lambda_mix}) for $\lambda_{dir}$ as a function of 
$\lambda_{obs}$ for given values of $\lambda_{1}$, $\lambda_{2}$, \Rchic\ and 
\Rot\ yields:
\begin{equation}
    \label{eq_lambda_rel}
    \lambda_{dir}(\lambda_{obs}\, |\, \lambda_{1}, \lambda_{2}, \Rchic, \Rot )=\frac{a_{obs} \lambda_{obs} - a_1\, f_1\, \lambda_1 - a_2\, f_2\, \lambda_2}{a_{obs} - a_1\, f_1\, - a_2\, f_2 }.
\end{equation}
In this equation \Rchic\ and \Rot\ enter via the fractions $f_i$.
 On the other hand, as both depend also on $\lambda_{dir}$, an
iterative procedure is applied starting with $\lambda_{dir}=0$. 

\subsubsection{Polarisation dependence of the efficiencies} 

For each tested pure \chic\ polarisation state with the corresponding set of 
values 
$\lambda_{dir},\ \lambda_1,\ \lambda_2$, new efficiencies \epsg\ and \epsjp\ 
are determined.

\paragraph{\jpsi\ efficiencies:} Assuming no
dependence of $\lambda$ on $x_F^{\jpsi}$ and $p_T^{\jpsi}$ 
(approximately valid within
the uncertainty of our measurement~\cite{roberto_jpsi}), we can write 
$\epsjp(\lambda)$ as:
\begin{displaymath}
\varepsilon_{J/\psi}(\lambda)=\frac{N_{reco}^{\jpsi}}{N_{gen}^{\jpsi}}
\end{displaymath}
\begin{equation}
\label{eq_efficiency_1}
 =\frac{\int A(\cos\theta', P)\cdot M(P)\cdot(1+\lambda\cos^2\theta')\cdot d\cos\theta' dP}{\int M(P)\cdot(1+\lambda\cos^2\theta')\cdot d\cos\theta' dP},
\end{equation}
where $\theta'$ is the polar angle in the polarisation frame, $P$ is shorthand 
for all the other phase space variables, $A(\cos\theta', P)$ is the acceptance 
at the kinematical point $(\cos\theta', P)$, $M(P)$ is the squared matrix 
element in $P$ and $\lambda$ is the polarisation parameter. 
After calculating the integrals we find:
\begin{equation}
\label{eq_efficiency_2} 
\varepsilon_{J/\psi}(\lambda)=\varepsilon_{J/\psi}(\lambda=0)\frac{1+\lambda\cdot\langle\cos^2\theta'\rangle}{(1+\lambda/3)},
\end{equation}
where $\langle\cos^{2}\theta'\rangle$ is given by:
\begin{displaymath}
    \langle\cos^{2}\theta'\rangle=\frac{\int A(\cos\theta', P)\cdot M(P)\cdot\cos^{2}\theta'\cdot  d\cos\theta' dP}{\int A(\cos\theta', P)\cdot M(P)\cdot  d\cos\theta' dP}.
\end{displaymath}
All \jpsi\ efficiencies, both for direct \jpsi\ and for \jpsi\ from the two 
\chic\ states, are calculated using Eq.~\ref{eq_efficiency_2}.

\paragraph{Photon efficiencies:} To determine the effect of polarisation on 
the photon efficiencies we start with the formula:
\begin{equation}
 \label{eq_efficiency_3}
 \epsgj = \frac{\nxcjm}{\njpxcjm},
\end{equation}
where $J,\, M$ denotes the polarisation state considered,
\njpxcjm\ is the number of \jpsi\ coming from $\chi_{cJ}$ and
\nxcjm\ is the number of observed \chicj. The value of
\njpxcjm\ is obtained from a fit of the $l^+l^-$ mass distribution, where each 
event enters with a weight:
\begin{equation}
 \label{eq_weight_1}
 w(\cos\theta', \lambda_{J,M})=\frac{1+\lambda_{J,M} \cdot\cos^{2}\theta'}{(1+\lambda_{J,M}/3)}.
\end{equation}

The value of \nxcjm\ is obtained from a fit of the \Dm\ distribution of true 
\chicj\ (using MC generator information  to select the correct \jpsi\gam\ 
combination), where the weight for each entry in the histogram
 corresponds to a certain pure polarisation state of \chicj, calculated by 
Eq.~\ref{eq_w_t_k}.

\section{Systematic Uncertainties}\label{systematics}
\subsection{Uncertainties from reconstruction, calibration, simulation and
background subtraction}\label{systematics:RecoCalSimBg}

With the exception of \jpsi\ counting, all of the systematic uncertainties in
the measurement of \Rchic\ are common to the \ee\ and \mm\ channels. The
\jpsi\ counting systematic uncertainties are estimated to be 2\% in the
electron channel and 0.25\% in the muon channel.\\
The remaining systematic uncertainty estimates include:

\subparagraph{\bf \chic\ counting:}  

\begin{itemize}
\item[-] photon selection ($7\%$);
\item[-] variation of \lplm\ mass window (2\%);
\item[-] \chic\ counting procedure ($4\%$) including:
  \begin{itemize}
  \item[$\cdot$] variation of the fixed parameters of the double-gaussian fit:
        $\Dm_{\chico}$, $\sigma_{\chico}$, $\Delta M(\chict)-\Delta M(\chico)$
        and $\frac{\sigma_{\chict}}{\sigma_{\chico}}$;
  \item[$\cdot$] fit with free $\Dm_{\chico}$ and/or $\sigma_{\chico}$;
  \item[$\cdot$] change of binning of the \Dm\ spectrum.
  \end{itemize}
\item[-]  Extensive tests were performed on the background determination with
the  mixed event procedure:
  \begin{itemize}
  \item[$\cdot$] variation of corrections corresponding to combinations of
       \jpsi 's  with photons from \chic\ decays in ME which do not occur in
       SE ($+ 3\%$);
  \item[$\cdot$] relaxing the requirement of the same neutral cluster
       multiplicity in ME and SE ($\pm 2\%$);
  \item[$\cdot$] variation of the cut on the neutral cluster direction in ME
       with respect to SE ($\pm 3\%$);
  \item[$\cdot$] allowing for an additional, polynomial term in the background
       to improve the fit around the \chic\ signal yields an asymmetric
       uncertainty ($+4\%$).
  \end{itemize}
  The total contribution from the background description to the systematic
  uncertainty in the \chic\ counting is estimated to range between $-4\%$
  and $+6\%$.
\end{itemize}

\subparagraph{\bf Efficiency evaluation}   

\begin{itemize}
\item[-] the use of different kinematic distributions for the generation of
         the \jpsi\ (see Ref.~\cite{hbpsiprime} and~\cite{HB_bbar}), affecting
         both $\frac{\epsxc}{\epsdir}$ and \epsg, introduces a systematic
         effect on \Rchic\ of $4\%$;
\item[-] tests on the photon efficiency simulation were performed including
         the comparison between real data and Monte Carlo of the \gam\
         conversion yield and of the detection efficiency of photons
         from electron Bremsstrahlung. The overall systematic uncertainty on
         \epsg\ determination and correction is found to be $6.5\%$.
\end{itemize}

The overall systematic uncertainty on \Rchic, evaluated as the quadratic sum
of the above terms, is therefore $^{+13}_{-12}\%$ for both \jpsi\ decay
channels.

The systematic uncertainty on \Rot\ is completely dominated by the
accuracy of the ECAL energy calibration which affects $\Delta M(\chico)$
which in turn affects the ratio \chicochict . A fine tuning of the ECAL
calibration as a function of the photon energy was performed using
the $\pinot \to \gam \gam$ signal. An absolute calibration accuracy of
$\sim 2\%$ was obtained and on $\Dm_{\chico}$ of $\sim 8~\mmev$. By scanning
$\Delta M(\chico)$ in such range around the nominal
value~\cite{PDG}, a variation of \chicochict\ (and thus \Rot) of $35\%$ is
obtained. No effect on \Rot\ is observed by changing the other fixed fit
parameters.

\subsection{Polarisation effects}\label{systematics:polarisation}

Since the direct \jpsi\ and \chic\ polarisations cannot be determined
separately from our data, we estimate instead systematic uncertainties  on the
reference values reported in Table~\ref{tab:results:res} (and denoted by
$\Rchic^{ref}$ and $\Rot^{ref}$ in the following) which were obtained with
the assumption of zero polarisation. The results of this study are expressed
as overall shifts of the values of \Rchic\ and \Rot\ due to the
average
polarisation
of directly produced \jpsi\ with uncertainties obtained from the maximum
variation of \chic\ polarisations allowed by the measurement:
\begin{equation}
    \label{eq_results_system}
    \begin{array}{lcl}
\frac{R_{\chi_{c}}-R_{\chi_{c}}^{ref}}{R_{\chi_{c}}^{ref}}=
+9.5\%^{+11\%}_{-7\%}\\[5mm]
\frac{\Rot-\Rot^{ref}}{\Rot^{ref}}=+0\%^{+16\%}_{-11\%},
    \end{array}
\end{equation}
where the following ingredients are used:
\begin{itemize}

\item The central values are obtained from the average measured value for
$\lambda_{obs}$ and with the assumption of no polarisation of \chico\ and
\chict\ ($\lambda_{obs}=-0.35$, $\lambda_{1}=0$ and $\lambda_{2}=0$, yielding
$\lambda_{dir}=-0.424$). Therefore, if the observed \jpsi\ polarisation were
due exclusively to direct \jpsi\  polarisation, the measured \Rchic\ would be
shifted up by $9.5\%$, while obviously no effect on \Rot\ is produced.

\item The variation bands in Eq.~\ref{eq_results_system} are obtained by
taking the extreme positive and negative variations of the central values
defined above, of all combinations of $\lambda_{obs}$
(varied in a $95\%$ c.l. range around the measured value, see
Eq.~\ref{eq:lambda_obs}) with $\lambda_{1}$ and
$\lambda_{2}$ (corresponding to the different pure helicity states M1 and M2):
\begin{itemize}
\item upper value: $\lambda_{obs}=-0.44$; $\lambda_{1}=-0.24$,
$\lambda_{2}=0.18$ for \Rchic; $\lambda_{1}=-0.24$,
$\lambda_{2}=-0.18$ for \Rot;
\item lower value: $\lambda_{obs}=-0.26$; $\lambda_{1}=0.22$ and
$\lambda_{2}=-0.18$ for \Rchic; $\lambda_{1}=0.22$ and
$\lambda_{2}=0.18$ for \Rot.
 \end{itemize}

\item Different polarisation values give overlapping ranges of possible 
\Rchic\ and \Rot\ values. Any value in each range is equally probable.  
Thus, even if the error on $\lambda_{obs}$ was Gaussian distributed, the 
errors of \Rchic\ and \Rot\ would not be Gaussian distributed. 
To take into account that the polarisation parameter $\lambda_{obs}$ was 
determined as an average over the whole accepted phase space and over 
different materials, $\lambda_{obs}$ was varied in a $\pm 2\sigma$ range with 
equal weights. Selecting the maximum deviations the measured values \Rchic and 
\Rot\ would have to be scaled:
\begin{equation}
    \label{eq_new_results_system}
    \begin{array}{lcl}
R_{\chi_{c}} = f_{\Rchic} \cdot R_{\chi_{c}}^{ref}\\[5mm]
\Rot = f_{\Rot} \cdot \Rot^{ref},
    \end{array}
\end{equation}
with $f_{\Rchic}\in$ [1.02, 1.21] and $f_{\Rot}\in$ [0.89, 1.16], 
where the uncertainties due to polarization are fully contained in the 
ranges given.

\item Note that the correlation between the values of \Rchic\ and \Rot\ are 
ignored in Eqs. \ref{eq_results_system} and \ref{eq_new_results_system}.
\end{itemize}

\section{Results}\label{results}
\subsection{\Rchic}\label{rxc}

The measured values for \Rchic\ are computed from Eq.~\ref{eq:method:rxc}, 
assuming zero \jpsi\ polarisation and are reported in 
Tab.~\ref{tab:results:res}, 
separately for muon and electron channels and combined sample. When averaged 
over decay channels and target materials, a value of 
\begin{equation}
\Rchic = 0.188\pm0.013_{st}{^{+0.024}_{-0.022}}_{sys}
\label{eq:results:rxc}
\end{equation}
is obtained. The quoted uncertainties include all systematic contributions
(except the polarisation contribution which is given in 
Eq.~\ref{eq_results_system} as a variation band at $95\%$ c.l). 
The following observations can be made:
\begin{itemize}
\item The results obtained in the two lepton channels are compatible within 
$1\sigma$ in both C and W samples. No measurement for the Ti in the electron 
channel is possible due to the low statistics;
\item The values of \Rchic\ obtained separately in the three target samples 
are consistent with each other;
\item The present result is lower than most values published in the literature
in pN interactions (see Tab.~\ref{tab:production:exper-res} and 
Fig.~\ref{fig:production:rchic-proton}).
Despite the fact that the various available measurements are taken at widely 
differing centre of mass energies, they are for the most part compatible 
within $\sim 1.5\sigma$, except for E705 ($2.3\sigma$) and ISR 
($3.3\sigma$).\\ 
The present measurement is lower than the previous \hb\ result~\cite{xcherab}
by about $2\sigma$. The two analyses are quite similar, although more 
extensive systematic checks have been performed in connection with the present 
one. These checks did not uncover any error in the previous analysis and we 
thus believe that the differences are largely statistical. 
The average of the two \hb\ results, $\Rchic = 0.198^{+0.028}_{-0.026}$,
differs by less than $1\sigma$ from the result of Eq.~\ref{eq:results:rxc}.
\end{itemize}
\begin{table*}[htb]
\begin{center}
\begin{tabular}{c c c c}
\hline
\hline
Mat. & \ee\ & \mm\ & combined \\
\hline
\multicolumn{4}{c}{ \Rchic\ } \\
\hline
C & $0.174\pm0.029_{st}{^{+0.022}_{-0.021}}_{sys}$ & $0.190\pm0.018_{st}{^{+0.024}_{-0.022}}_{sys}$&$0.185\pm0.015_{st}{^{+0.024}_{-0.022}}_{sys}$\\
Ti &          -                                    & $0.197\pm0.079_{st}{^{+0.025}_{-0.023}}_{sys}$&$0.197\pm0.079_{st}{^{+0.025}_{-0.023}}_{sys}$\\
W & $0.202\pm0.055_{st}{^{+0.026}_{-0.024}}_{sys}$ & $0.191\pm0.034_{st}{^{+0.025}_{-0.022}}_{sys}$&$0.194\pm0.029_{st}{^{+0.025}_{-0.023}}_{sys}$\\
Tot &$0.180\pm0.025_{st}{^{+0.023}_{-0.021}}_{sys}$& $0.190\pm0.015_{st}{^{+0.024}_{-0.022}}_{sys}$&$0.188\pm0.013_{st}{^{+0.024}_{-0.022}}_{sys}$\\
\hline
\hline
\multicolumn{4}{c}{ \Rot\ } \\
\hline
C  &$0.82\pm0.32_{st}$ &$1.23\pm0.27_{st}$ & $1.06\pm0.21_{st}\pm0.37_{sys}$ \\
Ti &     -             &$0.67\pm0.67_{st}$ & $0.67\pm0.67_{st}\pm0.23_{sys}$ \\
W  &$0.73\pm0.49_{st}$ &$1.27\pm0.53_{st}$ & $0.98\pm0.36_{st}\pm0.34_{sys}$ \\
Tot&$0.79\pm0.27_{st}$ &$1.17\pm0.22_{st}$ & $1.02\pm0.17_{st}\pm0.36_{sys}$ \\
\hline
\hline
\multicolumn{4}{c}{ $\frac{\sigma(\chico)}{\sigma(\chict)}$ } \\
\hline
C  &$0.47\pm0.19_{st}$ &$0.70\pm0.16_{st}$ & $0.60\pm0.12_{st}\pm0.21_{sys}$ \\
Ti &     -             &$0.38\pm0.38_{st}$ & $0.38\pm0.38_{st}\pm0.13_{sys}$ \\
W  &$0.41\pm0.28_{st}$ &$0.72\pm0.30_{st}$ & $0.56\pm0.21_{st}\pm0.20_{sys}$ \\
Tot&$0.45\pm0.16_{st}$ &$0.66\pm0.13_{st}$ & $0.57\pm0.10_{st}\pm0.20_{sys}$ \\
\hline
\hline
\end{tabular}
\caption{\it Measured values of \Rchic, \Rot\ and
$\frac{\sigma(\chico)}{\sigma(\chict)}$ in \ee, \mm\ and combined sample
for the different materials and the full data sample.}
\label{tab:results:res} 
\end{center}
\end{table*}

\subsection{\Rot}\label{r12}

The measured values of \Rot\ are evaluated using Eq.~\ref{eq:method:R12}, 
assuming no polarisation for either the directly produced \jpsi 's or the 
\chic 's, and are summarised in Table \ref{tab:results:res}.

As above, no dependence on target material is observed. The results from the 
electron channel are consistently lower than the muon results, but nonetheless 
in agreement to within $1\sigma$ of the statistical uncertainties.

The final result averaged over decay channel and target material is:
\begin{equation}\label{eq:results:r12}
\Rot = 1.02\pm0.17_{st}\pm0.36_{sys}
\end{equation}
where the systematic uncertainty does not include the polarisation 
contribution which is given in Eq.~\ref{eq_results_system} as a variation band 
at $95\%$ c.l. The \jpsi\ yields from \chico\ and  \chict\ are therefore found 
to be equal, although with large uncertainties.

\subsection{Dependence on kinematic variables}\label{kinematics}

A study of the dependence of \Rchic\ on the kinematic variables $x_F^{\jpsi}$ 
and $p_T^{\jpsi}$ in the ranges covered by \hb\ ($x_F^{\jpsi} \in 
[-0.35,0.15]$, $p_T^{\jpsi} \lesssim 5~GeV/c$) was performed by 
applying the described procedure in five $x_F^{J/\psi}$ and three 
$p_T^{\jpsi}$ intervals respectively. The resulting
distributions, for both channels combined, are shown in 
Fig.~\ref{fig:results:rchic_vs_kin}a)-b). The data is compatible with a flat 
dependence of \Rchic\ on both kinematic variables, although more complex 
dependences cannot be ruled out.
\begin{figure*}[htb]
\resizebox{1.0\textwidth}{!}{%
\epsfig{file=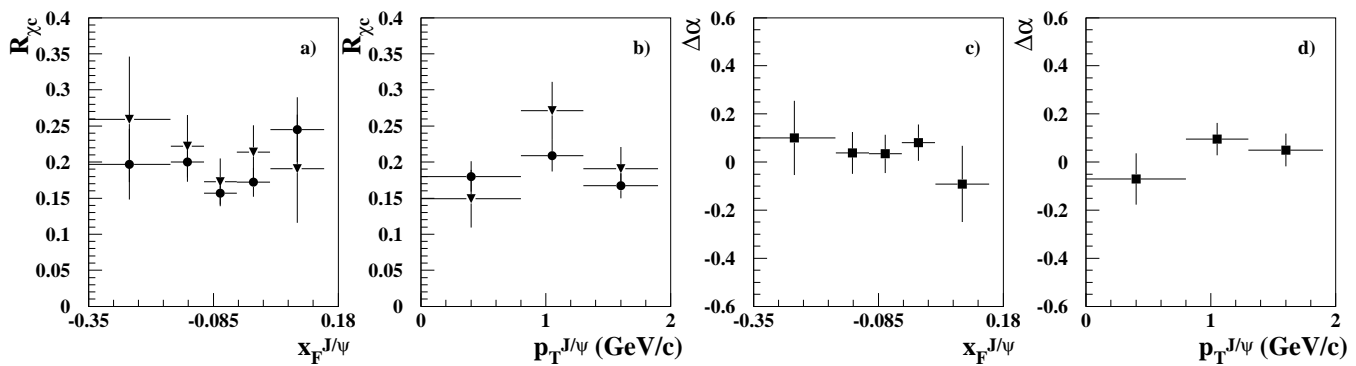}
}
\caption{Dependence of \Rchic\ on $x_F^{\jpsi}$ (a) and $p_T^{\jpsi}$ (b)
(circles: C; triangles: W). Dependence of 
$\Delta \alpha = \alpha_{\chic}-\alpha_{\jpsi}$ on $x_F^{\jpsi}$ (c) and 
$p_T^{\jpsi}$ (d). Only statistical errors are shown.}
\myfiglabel{fig:results:rchic_vs_kin}
\end{figure*}

\subsection{A-dependence}\label{Adep}

The atomic mass number (A) dependence of inclusive cross sections is often 
parameterised as a power law:
\begin{equation}
        \sigma_{pA} = \sigma_{pN} A^\alpha
\end{equation}
where $\sigma_{pA}$ is the inclusive production cross section in collisions of 
protons with a nuclear target of atomic mass number A, $\sigma_{pN}$ is the 
average cross section in collisions of protons with a single nucleon and 
$\alpha$ characterises the A dependence of the cross section. The difference 
between $\alpha$ for \jpsi\ production and that for \chic\ production can be 
computed from the measured values of $R_{\chi_c}$ for C and W targets given in 
Tab.~\ref{tab:results:res} from the following formula:
\begin{equation}\label{eq:results:rchic_material_xf}
\Delta \alpha = \alpha_{\chic}-\alpha_{\jpsi} = \frac {1.}{log \frac{A_W}{A_C}} \cdot log \frac{{\Rchic}^{W}}{{\Rchic}^{C}}
\end{equation}
where $A_W = 184$ and 
$A_C = 12$ are the tungsten and carbon atomic mass numbers. The results, 
plotted as a function of $x_F^{\jpsi}$ and $p_T^{\jpsi}$ are shown in 
Figs.~\ref{fig:results:rchic_vs_kin}c), d). Averaged over the visible 
$x_F^{\jpsi}$ and $p_T^{\jpsi}$ range, $\Delta \alpha = 0.05\pm 0.04$. 
The predictions of the various production models for $\Delta \alpha$ are all 
within the uncertainties of the measurement~\cite{ramonavogt}.

\subsection{\chic\ cross sections and ratio}\label{xsect}

From Eq.~\ref{eq:method:sigma12} we obtain the values for the cross
section ratio $\frac{\sigma(\chico)}{\sigma(\chict)}$
under the assumption of zero polarisation for both 
\jpsi\ and \chic. The results are reported in Tab.~\ref{tab:results:res}. The 
target material averaged result is:
\begin{equation}\label{eq:results:sig12}
\frac{\sigma(\chico)}{\sigma(\chict)} = 0.57\pm0.23
\end{equation}
where the uncertainty includes the systematic contributions (except 
polarisation - see above). The \chic\ production cross sections, defined as:
\begin{equation}
\sigma({\chic}_{i}) = \frac{\sigma(\jpsi) {\Rchic}_{i}}
{Br({\chic}_{i} \rightarrow \jpsi \gam)} , i=1,2
\label{eq:results:sigma12}
\end{equation}
are calculated using the estimate of the total \jpsi\ cross section at 
$\sqrt{s} = 41.6~GeV$,
$\sigma(\jpsi) = (502 \pm 44)$~nb/nucleon reported in \cite{ourcharmonium} and
assuming that \Rchic\ is independent of $x_F^{\jpsi}$ over the full 
$x_F^{\jpsi}$ and $p_T^{\jpsi}$ range. The following target material averaged 
values are obtained:
\begin{equation}\label{eq:results:sigxc}
\begin{array}{l}
\sigma(\chico) = 133\pm35~\mathrm{nb/nucleon}; \\
\sigma(\chict) = 231\pm61~\mathrm{nb/nucleon}.
\end{array}
\end{equation}
leading to a total \chic\ production cross section 
$\sigma(\chic) = 364\pm74~\mathrm{nb/nucleon}$.
Fig.~\ref{fig:production:rchic-proton} shows all available measurements of 
the \chico\ and \chict\ production cross sections and their ratio in 
proton-nucleus interactions at fixed-target energies. 

\section{Conclusion}\label{conclusions}
We have presented a new measurement of the fraction of all \jpsi\ mesons
produced through \chic\ decay (\Rchic), performed with the \hb\ detector in
pC, pTi and pW interactions at $920~GeV/c$ ($\sqrt{s}=41.6~GeV$).
The \chic\ mesons were detected in the \jpsi\gam\ decay mode, and the \jpsi\
in both \mm\ and \ee\ decay modes. The detector acceptance was flat in
$p_T^{\jpsi}$ and extended from $x_F^{\jpsi}=-0.35$ to $x_F^{\jpsi}=0.15$.

The measurement is based on a total sample of $\sim 15000$ \chic, the
largest ever observed in pA collisions.
Apart from lepton identification requirements, the analysis is identical for
the two channels. The separate results for the two channels are found to be
in agreement with each other in all respects.

The measured value $\Rchic = 0.188\pm0.013_{st}{^{+0.024}_{-0.022}}_{sys}$ is
$\sim 2\sigma$
lower than the previously published result from \hb. Our new value is also
lower than, but not incompatible with, most of the previously published values
obtained from pN interactions, independent of the centre of mass energies and
the kinematic ranges of the measurements. The present result supports the NRQCD
calculations \cite{ramonavogt}. When taken together with the already published
result of \hb\ on $\psi '$ production~\cite{hbpsiprime}, the fraction of all
\jpsi\ mesons coming from decays of higher mass charmonium states is found to
be $\sim 27\%$.

By separately counting the contribution of \chico\ and \chict, we obtain
a ratio of the two states $\Rot = {\Rchic}_1/{\Rchic}_2 = 1.02\pm0.40$ and a
cross section ratio $\frac{\sigma(\chico)}{\sigma(\chict)} = 0.57\pm0.23$.
The \chico\ and \chict\ cross sections are measured to be
$\sigma(\chico) = 133\pm35~nb$/nucleon and
$\sigma(\chict) = 231\pm61~nb$/nucleon in the full $x_F^{\jpsi}$ range.

No significant departure from a flat dependence of \Rchic\ on the kinematic
variables $ x_F^{\jpsi}$ and $p_T^{\jpsi}$ is found within the limited
accuracy of our measurement. No significant difference in the A-dependence of
\chic\ and \jpsi\ production is found within the limits of the available
statistics.

For the first time, an evaluation of the effect of polarisation of \jpsi\ and
\chic\ on the measured values of \Rchic\ and \Rot\ was performed. The
behaviour of \Rchic\ and \Rot\ as a function of the polarisation, expressed by
the $\lambda$ parameter, was studied with the conclusion that 
\Rchic\ and \Rot\ are uncertain with factors in the ranges [1.02,1.21] and 
[0.89,1.16], respectively, ignoring correlations between the two.

No mention of the influence of polarisation on the measurement of \Rchic\ can 
be found in any of the previous measurements. Nonetheless, we suspect that all 
measurements are subject to similar uncertainties to greater or lesser 
extents, depending on the geometry of the apparatus used.

\section{Acknowledgements}             
We express our gratitude to the DESY laboratory for the strong support in
setting up and running the HERA-B experiment. We are also indebted to the
DESY accelerator group for the continuous efforts to provide good and
stable beam conditions. 
The HERA-B experiment would not have been possible without the enormous
effort and commitment of our technical and administrative staff. It is a
pleasure to thank all of them. \\

\appendix
\section{$\chi_c$ angular distributions}

%
\begin{table*}[htb] 
\centering
\begin{tabular}{c c c c}
\hline
\hline
$_{\textstyle{\sigma}} \diagdown ^{\textstyle{\sigma'}}$  & -1 & 0 & 1\\
\hline
-1 & $\frac{1+\cos^{2}\theta'}{2}$ & $-\frac{\sin\theta'\cos\theta'}{\sqrt{2}}e^{i\phi'}$ & $\frac{\sin^{2}\theta'}{2}e^{2i\phi'}$\\
0 & $-\frac{\sin\theta'\cos\theta'}{\sqrt{2}}e^{-i\phi'}$ & $\sin^{2}\theta$ & $\frac{\sin\theta'\cos\theta'}{\sqrt{2}}e^{i\phi'}$\\ 
1 & $\frac{\sin^{2}\theta'}{2}e^{-2i\phi'}$ & $\frac{\sin\theta'\cos\theta'}{\sqrt{2}}e^{-i\phi'}$ & $\frac{1+\cos^{2}\theta'}{2}$\\
\hline
\hline
\end{tabular}   
\caption{Helicity density matrix for the $J/\psi$ decay as defined in 
(\ref{tab_density_matrix})}.
\label{tab_density_matrix}
%
\begin{center}
%
%
%
%
\begin{tabular}{r l c c c c }
\multicolumn{6}{c}{J=1} \\       
\hline 
\hline 
 &       &  \multicolumn{2}{c}{$K^{1,M}_i$ (general)} & \multicolumn{2}{c}{$K^{1,M}_i$ (E1 only)} \\[1mm]
\hline
 i &     $T^1_i$ & $M=0$ &   $M=1$   & $M=0$ & $M=1$ \\
\hline
 1 & 1 & $A^{2}_{1}$ & $\frac{1}{2} (A^{2}_{0}+ A^{2}_{1})$ & 0.5 & 0.5\\ 
 2 & $\cos^{2}\theta$ & $A^{2}_{0} -A^{2}_{1}$ &$ \frac{1}{2} (-A^{2}_{0} +A^{2}_{1})$ & 0 & 0\\
 3 & $\cos^{2}\theta^{\prime}$ & $-A^{2}_{1}$ & $\frac{1}{2} (A^{2}_{0} -A^{2}_{1})$  & -0.5 & 0\\
 4 & $\cos^2\theta\cos^{2}\theta^{\prime}$ & $A^{2}_{0} +A^{2}_{1}$&  $-\frac{1}{2} (A^{2}_{0} + A^{2}_{1})$ & 1.0 & -0.5\\
5 & $\sin2\theta\sin2\theta^{\prime}\cos\phi^{\prime}$ & $-\frac{1}{2}A_{0}A_{1}$ & $\frac{1}{4}A_{0}A_{1}$ & -0.25 & 0.125\\
\hline
\hline
\end{tabular}

\vspace{3mm}

\begin{tabular}{r l c c c c c c}
\multicolumn{8}{c}{J=2} \\       
\hline
\hline
 &       &  \multicolumn{3}{c}{$K^{2,M}_i$ (general)} & \multicolumn{3}{c}{$K^{2,M}_i$ (E1 only)} \\[1mm]
\hline
i &     $T^2_i$ & $M=0$ &   $M=1$   & $M=2$   & $M=0$ & $M=1$ & $M=2$\\
\hline
1 & 1 & $\frac{1}{4}A^2_0 + \frac{3}{8} A^2_2$ & $\frac{1}{2}A^2_1 + \frac{1}{4}A^2_2$ & $\frac{3}{8}A^2_0 + \frac{1}{2}A^2_1 + \frac{1}{16} A^2_2$ & 0.25 & 0.3 & 0.225  \\    
2 & $\cos^{2}\theta$ & $-\frac{3}{2}A^2_0 + 3 A^2_1 -\frac{3}{4}A^2_2$ & $\frac{3}{2}A^2_0 -\frac{3}{2}A^2_1$ & $-\frac{3}{4}A^2_0 + \frac{3}{8} A^2_2$ & 0.3 & -0.3 & 0.15\\
3 & $\cos^{4}\theta$ &  $  \frac{9}{4}A^2_0 -3 A^2_1  +  \frac{3}{8} A^2_2$ & $  -\frac{3}{2}A^2_0  +   2 A^2_1    -\frac{1}{4} A^2_2$ &  $\frac{3}{8}A^2_0 - \frac{1}{2}A^2_1  +  \frac{1}{16} A^2_2$ & -0.45 & 0.3 & -0.075 \\ 
4 & $\cos^{2}\theta^{\prime}$ & $  \frac{1}{4}A^2_0 + \frac{3}{8} A^2_2$ & $- \frac{1}{2}A^2_1+ \frac{1}{4}A^2_2$ & $  \frac{3}{8}A^2_0 - \frac{1}{2}A^2_1 + \frac{1}{16} A^2_2$ & 0.25 & 0 & -0.075\\ 
5 & $\cos^{2}\theta\cos^{2}\theta^{\prime}$ & $  -\frac{3}{2}A^2_0  -3   A^2_1   -\frac{3}{4}  A^2_2$ &  $  \frac{3}{2} A^2_0  +  \frac{3}{2}  A^2_1 $ &  $-\frac{3}{4}A^2_0    +  \frac{3}{8} A^2_2$ &  -1.5 & 0.6 & 0.15\\ 
                6 & $\cos^{4}\theta\cos^{2}\theta^{\prime}$ & 
                $  \frac{9}{4}A^2_0  +  3 A^2_1  +  \frac{3}{8} A^2_2$ & 
                $  -\frac{3}{2} A^2_0  - 2 A^2_1    -\frac{1}{4} A^2_2$ & 
                $ \frac{3}{8} A^2_0  +   \frac{1}{2}A^2_1  +   \frac{1}{16}A^2_2$ & 
         1.35 & -0.9 & 0.225\\ 

                7 & $\sin^{2}\theta^{\prime}\cos2\phi^{\prime}$ & 
                $-\frac{\sqrt{6}}{4}A_{0}A_{2}$ 
                & 0 & 
                $\frac{\sqrt{6}}{8}A_{0}A_{2}$ &
                 -0.15 & 0 & 0.075\\ 
                
                8 & $\cos^{2}\theta\sin^{2}\theta^{\prime}\cos2\phi^{\prime}$ &
                $\sqrt{6}A_{0}A_{2}$  &
                $-\frac{\sqrt{6}}{2}A_{0}A_{2}$ &
                 0 & 
                 0.6 & -0.3 & 0\\ 

                9 & $\cos^{4}\theta\sin^{2}\theta^{\prime}\cos2\phi^{\prime}$ & 
                $-\frac{3\sqrt{6}}{4} A_{0}A_{2}$  & 
                $\frac{\sqrt{6}}{2} A_{0}A_{2}$  &  
                $-\frac{\sqrt{6}}{8} A_{0}A_{2}$  & 
                -0.45 & 0.3 & -0.075\\ 
                
                10 & $\sin2\theta\sin2\theta^{\prime}\cos\phi^{\prime}$ & 
                $\frac{\sqrt{3}}{4}A_{0}A_{1}+ \frac{3\sqrt{2}}{8}A_{1}A_{2}$ &      
                $-\frac{\sqrt{3}}{4}A_{0}A_{1}$ &
                $\frac{\sqrt{3}}{8} A_{0}A_{1}-\frac{3\sqrt{2}}{16}A_{1}A_{2}$   & 
                0.3 & -0.075 & -0.075\\ 
                
                11 & $\cos^{2}\theta\sin2\theta\sin2\theta^{\prime}\cos\phi^{\prime}$ & $-\frac{3\sqrt{3}}{4}A_{0}A_{1}-\frac{3\sqrt{2}}{8}A_{1}A_{2}$ &  
                $\frac{\sqrt{3}}{2}A_{0}A_{1}+\frac{\sqrt{2}}{4}A_{1}A_{2}$ &  $-\frac{\sqrt{3}}{8}A_{0}A_{1}-\frac{\sqrt{2}}{16}A_{1}A_{2}$  & 
                -0.45 & 0.3 & -0.075\\ 
                
                \hline
                \hline
                \end{tabular}
                \end{center}
\caption{The angular distribution terms $T^J_i(\theta,\, \theta',\, \phi')$ and the coefficients $K^{J,M}_i(A_{|\nu|})$ as defined in 
 (\ref{eq_w_t_k_prime}) for $J=1,2$. The last columns give the numerical values for the coefficients $K^{J,M}_i$ for different $M$ with the assumption that only the electric dipole transition contributes (see (\ref{eq_eone_restriction})).}         \label{tab_w_t_k_j}
\end{table*}

The angular decay distribution of a pure polarisation state $|J,M\rangle$ is given as an expansion into helicity amplitudes $A^J_{|\nu|}$ by \cite{Olsson}:
\begin{eqnarray}
        \label{eq_sumhel_chi_dfunc}
& &     W^{J,M}(\theta,\, \theta',\, \phi')=\\ \nonumber
& &      \sum_{\nu,\, \nu'=-J}^{+J} \sum_{\mu=\pm 1} d^{J}_{M\, \nu}(\theta)\ d^{J*}_{M\ \nu'}(\theta)\ \, A^J_{|\nu'|}\ \rho^{\sigma \sigma'}\!(\theta',\, \phi')
\end{eqnarray}
with the density matrix for the $J/\psi$ helicity (\ref{tab_density_matrix}):
\begin{equation}
        \label{eq_density_matrix}
        \rho^{\sigma \sigma'}\!(\theta',\, \phi') = \sum_{\kappa=\pm 1}
        D^{1}_{\sigma\, \kappa }(\phi', \theta', -\phi')\
        D^{1*}_{\sigma'\, \kappa }(\phi', \theta', -\phi').
\end{equation}

Using the notation of \cite{Olsson} the angular distribution can be decomposed into terms with trigonometric expressions $T^{J}_i(\theta,\, \theta',\, \phi')$ and coefficients $K^{J,M}_i(A^J_{|\nu|})$:
\begin{equation}
        \label{eq_w_t_k_prime}
        W^{J,M}(\theta,\, \theta',\, \phi')= 
        \sum_i K^{J,M}_i(\,A^{J}_{|\nu|})\ T^J_i(\theta,\, \theta',\, \phi') .
\end{equation}
 The $K^{J,M}_i(A^{J}_{|\nu|})$ and $T^J_i(\theta,\, \theta',\, \phi')$ are reported for $J=1,2$ in Table \ref{tab_w_t_k_j}. The normalisations are 
for the angular distributions

\begin{equation}
        \int W^{J,M}(\theta,\, \theta',\, \phi')\ d\cos{\theta}\,d\phi\, d\cos{\theta'}\, d\phi'\ =\ \begin{array}{cl}  \frac{64 \pi^2}{9} &  {\rm for}\ J=1\\[2mm]  \frac{64 \pi^2}{15} &  {\rm for}\ J=2 \end{array} 
\end{equation}
 
 The helicity amplitudes $A^J_{|\nu|}$ can be expanded in multipole amplitudes (E1, M2, E3), see for example  \cite{Olsson}. With the restriction to electric dipole transitions the helicity amplitudes become:
\begin{eqnarray}
\label{eq_eone_restriction}
\nonumber
        J=1: &  A_0=\sqrt{\frac{1}{2}},\ \ A_1=\sqrt{\frac{1}{2}} \\
        &  \\
        \nonumber
        J=2: & A_0=\sqrt{\frac{1}{10}}, \ \ A_1=\sqrt{\frac{3}{10}}, \ \ A_2=\sqrt{\frac{3}{5}}
\end{eqnarray}
The Table \ref{tab_w_t_k_j} reports also the coefficients $K^{J,M}_i$ calculated with these values for the helicity amplitudes.
Hence, with the restriction to the lowest multipole, the angular distributions of a $\chi_c$ decay for a given polarisation state $|J,M\rangle$ is fully determined (obviously, the relative contributions of different polarisation states are not fixed).

\end{document}